\documentclass[prb,twocolumn,amsmath,amssymb,floatfix,superscriptaddress]{revtex4-2}
\usepackage{graphicx} 
\usepackage{booktabs}
\usepackage[dvipsnames]{xcolor}
\usepackage[version=4]{mhchem}
\usepackage{xspace, bm, braket}
\usepackage{siunitx}
\usepackage{verbatim}
\usepackage{subfigure}
\usepackage[colorlinks, citecolor=blue, linkcolor=blue, urlcolor=blue, breaklinks]{hyperref}
\usepackage{hyperref}
\include{Braket}
\usepackage{dsfont}
\usepackage[normalem]{ulem}
\usepackage{mathtools}

\newcommand{\NNO}{\ce{NdNiO2}}
\newcommand{\CCO}{\ce{CaCuO2}}

\newcommand{\angstrom}{\textup{\AA}}

\begin{document}

\title{Dependence of DFT+DMFT Results on the Construction of the Correlated Orbitals} 

\author{Jonathan Karp}
\email{jk3986@columbia.edu}
\affiliation{Department of Applied Physics and Applied Math, Columbia University, New York, NY 10027, USA}

\author{Alexander Hampel}
\affiliation{Center for Computational Quantum Physics, Flatiron Institute, 162 5th Avenue, New York, NY 10010, USA}

\author{Andrew J. Millis}
\affiliation{Center for Computational Quantum Physics, Flatiron Institute, 162 5th Avenue, New York, NY 10010, USA}
\affiliation{Department of Physics, Columbia University, New York, NY 10027, USA}

\date{\today}

\begin{abstract}
The sensitivity of Density Functional Theory plus Dynamical Mean Field Theory calculations to different constructions of the correlated orbitals is investigated via a detailed comparison of results obtained for the quantum material  NdNiO$_2$ using different Wannier and projector methods to define the correlation problem.   Using the same interaction parameters  we find that  the different  methods produce different results for the orbital and band basis mass enhancements and for the orbital occupancies, with differing implications regarding the importance of multiorbital effects and charge transfer physics.  Using interaction parameters derived from cRPA enhances the difference in results. For the isostructural cuprate CaCuO$_2$, the different methods give quantitatively different mass enhancements but still result in the same qualitative physics. 
\end{abstract}

\maketitle

\section{Introduction}

The quantum many body problem is both complex and difficult. ``Difficulty" refers to   the combination of the exponential growth of Hilbert space with system size and quantum entanglement (in particular the Fermion sign problem) which render standard methods for dealing with large interacting  problems ineffective. ``Complexity" refers to the issues involved in formulating the many-body problem, in particular defining and computing the large number of basis functions and interaction parameters required to capture the chemical and structural effects that distinguish e.g. aluminum, a low transition temperature superconductor well described by conventional Migdal-Eliashberg theory, from e.g. La$_{2-x}$Sr$_x$CuO$_4$, a high transition temperature superconductor believed to have properties inconsistent with conventional Migdal-Eliashberg theory.

Some methods, such as Density Functional Theory, map the correlation problem onto a one body problem with a self-consistently determined potential. While in principle exact, these methods fail in practice for materials with strong electronic correlations.  Other approaches, such as the coupled cluster theories of quantum chemistry \cite{bartlett2007coupled}, treat the full complexity of  molecular systems but work well only for relatively small, relatively weakly correlated systems where the level of quantum mechanical difficulty is not large. Conversely,  Bethe ansatz approaches \cite{Bethe1931} provide an exact solution for model systems, thus fully taking into account the quantum mechanical difficulty while omitting the complexity needed to describe real materials. 

A complete treatment of the full quantum many-body problem in all its difficulty and complexity is not currently feasible. For most systems of interest, progress has come from the combination of reducing the complexity by ``downfolding" the full  problem to a much smaller and therefore more tractable one and managing the difficulty via an approximate  solution of the  quantum many body problem as defined in the downfolded subspace. Downfolding typically involves the identification of a subset  of single particle states which are then used to construct the many-particle Fock space in which the quantum many body physics is to be solved and the projection of the Hamiltonian into this basis. A widely studied example of downfolding is the single band Hubbard model \cite{hubbard1963electron} in which the single-electron electronic structure is approximated as a one band tight binding model defined on a simple lattice and only the onsite term in the electron-electron interaction is retained.  

Interest in  downfolding has been renewed by the recent discovery \cite{Hwang2019} of superconductivity in infinite layer rare earth nickelates such as hole doped \NNO{}. This family of materials has  been of long-standing interest as a potential analog of the layered copper oxide (cuprate) superconductors \cite{Crespin1983reduced,hayward1999sodium, Anisimov99, Lee04}  and is currently the subject of intense theoretical and experimental research.  Electronic structure calculations \cite{Lee04, karp2020manybody, botana2019, nomura2019, hepting2019, wu2019, sakakibara2020model, gao2019electronic, zhang2019effective, jiang2019electronic, hirayama2019materials, Gu2020substantial, si2020topotactic, choi2019role, Liu2020electronic, olevano2020abinitio, leonov2020lifshitz, Kitatani2020nickelate} performed on \NNO{} reveal some similarities to the cuprates, including a similar nominal $d^9$ valence and a band of transition metal $d_{x^2-y^2}$ character crossing the Fermi level, but also some differences including the importance of Nd-derived bands, a rather different charge transfer energy, and potential relevance of other $d$-multiplet states. The similarities and differences raise the question: can one use essentially the same downfolded model to study superconductivity and other properties of the two compounds? 

The combination of Density Functional Theory and Dynamical Mean Field Theory (DFT+DMFT) \cite{Georges1996,Georges04,Kotliar06,Held06} has emerged as a powerful and widely used method for studying quantum materials, materials whose properties are determined by quantum many-body effects, because it combines a downfolding based on density functional theory that produces a reasonably realistic description of particular ``correlated" orbitals in the structural and chemical environment defined by the rest of the material with a many-body method that focuses on the solution of a local correlation problem.

The DFT+DMFT methodology has been broadly successful in describing the physics of many quantum materials \cite{Georges04,Kotliar06,Held06,Paul18}. Motivated by this success, many authors have performed  DFT+DMFT calculations on the infinite layer nickelates  \cite{karp2020manybody, karp2020comparative, lechermann2019late, wang2020hunds, kang2020infinitelayer, kang2020optical, leonov2020lifshitz, ryee2019induced, werner2019nickelate, petocchi2020normal, si2020topotactic, Kitatani2020nickelate, Gu2020substantial, liu2020doping}, but with different and sometimes conflicting results. Some papers find  that multiorbital physics is crucial to the correlation effects in \NNO{} \cite{lechermann2019late, lechermann2020multiorbital, wang2020hunds, kang2020infinitelayer, kang2020optical, petocchi2020normal}, while others \cite{karp2020manybody, Kitatani2020nickelate} claim that the important correlation physics lies in a single self-doped band. Some papers state that \NNO{} is in the Mott-Hubbard regime with little influence of charge transfer effects \cite{Kitatani2020nickelate, karp2020comparative}, while others claim that, as in the cuprates, charge transfer physics is important \cite{karp2020manybody}. Some of the differences arise from different choices of Coulomb interaction parameters, but it now appears that some of the differences arise from choices made in the downfolding procedure.

In the DFT+DMFT context, discussion of downfolding issues has centered on topics related to the appropriate treatment of interactions. Important questions have included the limits of applicability of the single-site dynamical mean field approximation, the question of which Coulomb matrix elements are treated dynamically and which via a mean field or one-loop theory \cite{biermann2003first, Werner_2016, Honerkamp2018, Nilsson2017} and the  ``double counting" problem of how the portion of the interactions included in the underlying density functional theory are accounted for ~\cite{liechtenstein1995fll, aichhorn2011importance, Held2007Electronic, czyzyk1994amf, park2014computing, Haule2015exact}. The issue of  abstracting a one electron basis for the correlated subspace  out of  a more chemically realistic background electronic structure has been assumed to be less problematic. 

In  this paper we show, using NdNiO$_2$ as an example,  that the aspect of downfolding involving the choice of basis set requires more attention than has heretofore been assumed. Different methods of downfolding, all of which reproduce the underlying band structure, are shown to lead to markedly different results for many body properties of interest. We trace the origin of the differences back to different partitioning of the band theory electronic states into correlated and uncorrelated orbitals. The results suggest that the accuracy of different downfolding approximations should be revisited. Our specific results are derived in the context of the layered nickelates but the conclusions should be more generally valid.

The rest of this paper is organized as follows. In section ~\ref{Methods} we review the DFT+DMFT method, describe the different downfolding methods, and provide details on the calculations performed in this paper. In section \ref{sec:orb_cont} we compare the physical content of the localized orbitals constructed in the different downfolding methods. Section \ref{sec:dmft_results} presents our results from DFT+DMFT calculations with the different downfolding methods. Finally, we offer further analysis and concluding thoughts in section \ref{sec:discussion}. 

\section{Methods \label{Methods}}

\subsection{Theoretical Overview}

In DFT+DMFT the main object of interest is the one electron Green's function  
\begin{equation}
    \hat G(r,r^\prime;\omega)=\left(\omega \mathds{1} - \hat H_{ref}- \hat \Sigma(r,r^\prime;\omega)\right)^{-1}
    \label{Gdef}
    \end{equation}
with non-interacting reference Hamiltonian  $\hat H_{ref}$ taken to be the Kohn-Sham Hamiltonian resulting from the solution of the equations of density functional theory and the self energy $\hat{\Sigma}$ constructed by identifying (on physical grounds) particular correlated orbitals with wave functions $\phi_\alpha^m(r-R^\alpha)$ corresponding to orbitals $m$ and localized near sites $R^\alpha$. The Green's function is calculated by making the single site DMFT approximation   
in which only the site-local matrix elements of the self energy between correlated orbitals are retained.  For example, in a transition metal oxide the $\phi_\alpha^m$ might be chosen to represent the orbitals in the 3$d$ shell of the transition metal ion with nucleus at site $R^\alpha$.  

The site-local self energy $\hat\Sigma^\alpha_{QI}$ is  obtained from the solution of a quantum impurity model, a $0$ space $+1$ time dimensional quantum field theory describing $m$ correlated orbitals coupled to a non-interacting bath. The interactions of the quantum impurity model  are chosen to represent the matrix elements of the screened Coulomb interaction among the correlated orbitals  $\phi_\alpha^m$. The one-electron parameters of  the quantum impurity model are determined from  a self-consistency equation relating the Green's function of the quantum impurity model $\hat G_{QI}$ to the projection onto site $\alpha$ of the full lattice Green's function. 

The specification of the correlated orbitals  and of their coupling to the other degrees of freedom in the solid is thus fundamental to the DFT+DMFT method. Two closely related methods, referred to in the literature as ``projector" and ``Wannier" methods, are widely used for this purpose. 
 
 In the projector methodology \cite{anisimov2005full, amadon2008plane}, outlined in \cite{aichhorn2009},  one predefines a set of atomic-like correlated orbitals $\ket{\tilde \phi_m^{\alpha}}$, typically chosen to be centered on positions $R^\alpha$ of particular atoms of interest with the symmetry appropriate to the correlated orbital of interest (e.g. transition metal $d$), and vanishing for $|r-R^\alpha|$ greater than some pre-set value. One then represents $\Sigma$ in  the ``Kohn-Sham''  basis of eigenstates $\psi_{\nu k}(r)$ of $H_{ref}$. All practical calculations retain only a finite set of bands within a window $\mathcal{W}$ (which may depend on $k$) so the  orbitals are defined as
  \begin{equation}
        \ket{\tilde{\phi}^{m}_\alpha} =\sum_{\nu,k\in \mathcal{W}(k)} \tilde P_{\nu,k}^{\alpha, m}\ket{\psi_{\nu,k}}
        \label{eq:Pdef1}
    \end{equation}
    with 
    \begin{equation}
        \tilde P_{\nu,k}^{\alpha, m}=\braket{\tilde{\phi}^{m}_\alpha | \psi_{\nu,k}}
    \end{equation}
  The $\ket{\tilde{\phi}^{m}_\alpha}$ defined in Eq.~\ref{eq:Pdef1}  are a sum over an incomplete basis and must be orthonormalized. The  result after orthonomalization is a set of states $\ket{\phi^m_\alpha}$ that deviate to some degree from the originally defined atomic like states $\ket{\tilde{\phi}_m^\alpha}$, and in particular have tails that extend outside the originally defined state radius. This consideration suggests that it is often advantageous to formulate the problem in as wide an energy range as feasible, to make the $\ket{\phi_m^\alpha}$ as similar as possible to the $\ket{\tilde\phi_m^{\alpha}}$.
  
  With the $\ket{\phi_\alpha^m}$ in hand one ``upfolds'' the self energy to the Kohn Sham basis  via  
 \begin{equation}
    \Sigma_{\nu\nu^\prime}(k,\omega)= \sum_{\alpha m m'} P_{\nu m}^{\alpha k} \left[\Sigma_{\text{QI}}^\alpha (\omega) \right]_{mm'}\left(P_{m' \nu^\prime }^{\alpha k}\right)^\star
    \label{sigmaupfold}
    \end{equation}
    where the $P$ are the coefficients in an expansion of $\ket{\phi^{m}_\alpha}$ in the $\ket{\psi_{\nu,k}}$:
     \begin{equation}
        P_{\nu m}^{\alpha k}=\braket{\psi_{\nu,k}|\phi^{m}_\alpha}
        \label{eq:Pdef}
    \end{equation}
    so that  full Green's function, Eq.~\ref{Gdef} is written
\begin{equation}
    \left[G^{latt}(k,\omega)\right]_{\nu\nu^\prime} =\left[\omega \mathds{1}- \hat H_{ref}(k)- \hat \Sigma(k,\omega)\right]^{-1}_{\nu \nu'}
    \label{eq:Glatt}
    \end{equation}
    The same basis transformation may be used to ``downfold" the lattice  Green's function to the basis of correlated orbitals, yielding the self-consistency equation relating the quantum impurity model Green's function $G_{QI}$ to the downfolded lattice Green's function:
    \begin{eqnarray}
G^{mm^\prime}_{QI;\alpha}(\omega)=\sum_{\nu,\nu^\prime,k}\left(P_{m \nu }^{\alpha k}\right)^\star G_{\nu,\nu^\prime}(k,\omega)P_{\nu^\prime m'}^{\alpha k }
\label{eq:SCE1}
\end{eqnarray}

Eq.~\ref{eq:SCE1} can be rearranged to determine the one-electron parameters of the quantum impurity model. It is important to note that the equation is formulated directly in terms of the projection of the lattice Green's function onto the pre-specified correlated orbitals and the upfolding of the impurity model self energy to the Kohn-Sham basis.  While it is possible to define the projection of the Kohn-Sham Hamiltonian onto the correlated orbitals, this Hamiltonian is not used in the formalism; in particular it is not  what enters the impurity model.

In the Wannier methodology one first identifies a set of $N$ Kohn Sham bands that are not significantly entangled with other energy bands. If the $N$ bands are contained within an energy window that does not contain any other bands, the identification is straightforward. In the more common case in which there is no energy window that fully isolates a relevant set of bands, a disentanglement procedure ~\cite{MLWF2} is performed in which an energy window $\mathcal{W}$ containing more than $N$ bands is defined and then at each $k$ point $N$ optimized bands are constructed as linear combinations of bands inside the entanglement window via

\begin{align}
    \ket{\psi_{\mu k}^\text{opt}} = \sum_{\nu \in \mathcal{W}} T_{\mu\nu}^{\text{dis}(k)} \ket{\psi_{\nu k}}
\end{align}
Following Ref. ~\cite{MLWF2} the disentangling  transformation $T_{\mu\nu}^{\text{dis}(k)} $ (which may be non-unitary) is chosen to  minimize a spread function that  ensures $k$-point connectivity, or ``global smoothness of connection''~\cite{MLWF2} in the optimized states and also includes an orthonormalization  step.

Finally,  $N$ Wannier states $\phi_a^I(r)$  localized at positions $R_a^I$ in  unit cell $I$ are defined as   \cite{MLWF2}.
\begin{equation}
    \phi_a^I(r)=\sum_{k,\nu}U_k^{a\nu}\ket{\psi_{\nu,k}^\text{opt}(r)} e^{-ik(r-R_a^I)}
    \label{Wannier}
\end{equation}
and a number $N_c\leq N$ of these  are designated as correlated orbitals.   

 The unitary operators ${U}^{m \nu}_k$ are chosen to optimize some desired property of the Wannier functions, typically localization about the Wannier centers $R_a =\braket{\phi_a|r|\phi_a}$. In the Maximally Localized Wannier Function (MLWF) procedure \cite{MLWF1,MLWF2} one  minimizes the average over all Wannier functions of the mean square positional uncertainty $\sum_a\delta R_a^2\equiv \sum_a \braket{\phi_a(r)|(r-R_a)^2|\phi_a}$ (the unit cell index is suppressed here since the localization is the same for each cell). In the Selectively Localized Wannier Function (SLWF) procedure \cite{Wang14} one minimizes the spread only of the designated correlated orbitals, and also optimizes the center position and symmetry of these states, but the procedure is otherwise the same. 
 
 The projection of the Kohn-Sham Hamiltonian onto the Wannier orbitals defines an $N$ orbital tight binding model with Hamiltonian $H^{ab}_{IJ}=\braket{\phi_a^I|H_{KS}|\phi_b^J}$. An important test of the Wannierization procedure is that the eigenvalues of $H^{ab}_{IJ}$ reproduce the Kohn-Sham bands with high precision: failure to reproduce the DFT band structure means that the Kohn-Sham Hamiltonian has matrix elements between the $\ket{\phi_a}$ and Bloch functions $\ket{\psi_{n k}}$ not included in Eq.~\ref{Wannier}, so that the Wannier basis is not a complete expression of the single particle physics in the relevant energy range.
 
 In the Wannier method the dynamical mean field self consistency is expressed in terms of the Wannier Green's function  
 \begin{equation}
     G^{ab}_{IJ}(\omega)=\left[\omega \mathds{1}-\hat H-\hat{\Sigma}(\omega)\delta_{IJ}\right]^{-1}_{IJab}
     \label{GWannier}
     \end{equation}
     where the self energy matrix has nonzero elements only in the $N_c\times N_c$ correlated orbital subspace, with these matrix elements being precisely equal to the quantum impurity model self energy. The Wannier analog of Eq. ~\ref{eq:SCE1} is then given by equating the quantum impurity model Green's function to the sub-block of the onsite $G$:
     \begin{equation}
         G_{QI}^{\tilde{a}\tilde{b}}(\omega)=G^{\tilde{a}\in N_c\tilde{b}\in N_c}_{II}(\omega)
         \label{SCE2}
     \end{equation}

In the Wannier method the self consistency is thus carried out directly in the Wannier basis, with ``upfolding" to the Kohn Sham basis only required for reasons outside the scope of this paper such as computing the charge density required for the ``full charge self consistency" step of the DFT+DMFT procedure. 

The Wannier procedure requires construction of $H_{IJ}^{ab}$, which makes it in a sense less elegant than the projector method, but as will be seen, the form of $H_{IJ}^{ab}$ provides physical insight, and the orbitals and energies permit the use of cRPA methods for computing the interactions. 

Advantages of the projector method include the ability  to specify in an intuitively or chemical reasonable manner the shape and location of the correlated  orbitals (subject to the orthonormalization issues discussed above) and the avoidance of the multiparameter optimization required to construct the Wannier functions. Advantages of the Wannier procedure include a flexibility in determining the correlated orbital wave function (which the Wannier method will adapt, for example, to changes in lattice constant). Additionally, analysis of  the intermediate tight-binding model can provide physical insight and permits the use of cRPA methods for computing the interactions.

Both the Wannier and the projector methods involve a specification of the correlated orbital wave function, imposed {\em a priori} in the projector method and computed as part of the process in the Wannier approaches. The importance of the specification for the correlation physics may be seen by consideration of a simple two-site model of a $d$ orbital of energy $\varepsilon_d$, a ligand (``$p$'') orbital of energy $\varepsilon_p$, a hybridization $t_{pd}$, and a correlation term $Ud^\dagger_\uparrow d_\uparrow d^\dagger_\downarrow d_\downarrow$. The strength of the correlation effects depends on both $U$ and $(\varepsilon_d-\varepsilon_p)/t_{pd}$. On the noninteracting ($U=0$) level, the model has two levels with energy difference $\Delta E=\sqrt{\left(\varepsilon_d-\varepsilon_p\right)^2+4t_{pd}^2}$. We see that a range of $t_{pd}$ and $\varepsilon_d-\varepsilon_p$ can fit the same energy difference; pinning down the parameters requires additional information such as the $d$ content of the states. In close analogy, in the solid state case the Kohn-Sham eigenvalues and eigenfunctions do not by themselves  determine the energies of the correlated orbitals or their relation to the uncorrelated orbitals. The values of the analogs of $\varepsilon_p$, $\varepsilon_d$, and $t_{pd}$ can be read off directly from the Wannier Hamiltonian, and they can be inferred from the projector results. The crucial finding of our paper is that different projector and Wannier methods which have identical bands on a DFT level produce different results for these parameters, leading to strikingly different correlation physics.

\subsection{Computational details}
We perform one-shot DFT+DMFT calculations for NdNiO$_2$  using different combinations of downfolding approach and energy window. The four cases we consider are: 

\begin{enumerate}
    \item MLWF: We construct 13 Wannier functions, corresponding to the Ni-$3d$, O-$2p$, Nd-$5d_{z^2}$ and Nd-$5d_{xy}$ orbitals.  We use a disentanglement energy window of $\SI{-9.5}{eV}$ to $\SI{6.1}{eV}$ and a frozen window from $\SI{-9.5}{eV}$ to $\SI{1.4}{eV}$ and maximally localize the total spread of all 13 Wannier functions. 
    
    \item SLWF: We construct 13 Wannier functions, corresponding to the same orbitals as the MLWF case with the same disentanglement and frozen windows. However, in this case we only minimize the spread of the 5 Ni-$3d$ Wannier functions and ignore the spread of the rest. 
    
    \item Projectors in an energy window from $-\SI{10}{eV}$ to $\SI{10}{eV}$ around the Fermi energy. This guarantees that all relevant low energy states are included in the window, and projectors are quite localized. 
    
    \item Projectors in an energy window from $-\SI{10}{eV}$ to $\SI{3}{eV}$ around the Fermi energy, going high enough in energy to include the self-doping band but not the tail of the Nd-$5d_{z^2}$ and Nd-$5d_{xy}$ densities of states.

\end{enumerate}

For comparison we also used the two Wannier function methodologies to perform calculations for the ``infinite layer" high-T$_c$ cuprate CaCuO$_2$, retaining in this case only 11 bands because the Ca-$d$ states are far above the Fermi level. 

For the projector cases, we perform DFT calculations using WIEN2k~\cite{Blaha2018} and the  standard PBE GGA functional~\cite{PBE}. We use the experimental crystal structure with $a = b = \SI{3.92}{\angstrom}$ and $c = \SI{3.31}{\angstrom}$~\cite{Hwang2019} (\NNO{}) and   $a = b = \SI{3.86}{\angstrom}$ and $c = \SI{3.20}{\angstrom}$ (\CCO{}). We treat the Nd-$4f$ bands as core states. The DFT calculations are converged with an $RK_{max}=7$ and with a $k$-point grid of $40 \times 40 \times 40$. We use the dmftproj software~\cite{TRIQS/DFTTOOLS} to create the projectors. 

For the MLWF and SLWF cases we perform the DFT calculations with Quantum Espresso \cite{QE}. We use the same structure parameters and PBE functional as with Wien2k. We use PAW pseudopotentials with the the Nd-$f$ states in the core. We use a $k$-point mesh of $16 \times 16 \times 16$, an energy cutoff of $\SI{70}{Ry}$ for the wave functions, and an energy cutoff of $\SI{280}{Ry}$ for the density and potential. We use Wannier90 \cite{wannier90_v3} to create the Wannier functions. We find that wannierization results do not depend on the DFT code; Wien2k (using the wien2wannier program \cite{wien2wannier}) and Quantum Espresso give the same results.

We perform single-site DMFT calculations using the TRIQS software library~\cite{TRIQS, TRIQS/DFTTOOLS}. For \NNO{} we consider two cases: a ``two orbital"  theory treating the dynamical correlations among the two Ni-$e_g$ orbitals and a ``one orbital" theory treating only the $d_{x^2-y^2}$ orbital as correlated; for \CCO{} we perform a ``two orbital" calculation. The calculations are one-shot in the sense that the DFT density is not further updated. For the projector cases we also run calculations with full charge self consistency, finding that imposing full charge self consistency  does not alter the results significantly.

For the impurity problem, we choose the interactions to be of the Kanamori form \cite{Kanamori1963}:
\begin{multline}
    H = U \sum_m n_{m \uparrow}n_{m \downarrow} + \sum_{m<m', \sigma}[U'n_{m\sigma}n_{m'\Bar \sigma} \\
        + (U' - J)n_{m\sigma}n_{m'\sigma} 
        - Jc^\dagger_{m\sigma} c_{m\Bar \sigma} c^\dagger_{m' \Bar \sigma} c_{m' \sigma}] \\
        - J \sum_{m<m'} [c^\dagger_{m\uparrow}c^\dagger_{m\downarrow}c_{m'\uparrow}c_{m'\downarrow} + H.c.]
\end{multline}
with $U' = U - 2J$ and $\sigma \in \{\uparrow, \downarrow \}$ refers to the spin projection. Unless otherwise noted, we use an onsite Hubbard interaction of $U = \SI{7}{eV}$ and a Hund's coupling of $J = \SI{0.7}{eV}$, considered reasonable values for nickelates \cite{Nowadnick15}, and we use a temperature of $T = \SI{290}{K}$. We solve the impurity problem using CTHYB \cite{TRIQS/CTHYB}. We use Held's double counting formula~\cite{Held2007Electronic}:
 \begin{equation}
     \Sigma_{dc} = \frac{U + (D-1)(U-2J) + (D-1)(U-3J)}{2D-1} (n - 0.5),
 \end{equation}
where $D$ is the number of correlated orbitals and $n$ is the density of the correlated orbitals obtained from their local non-interacting Green's function. We employ the maximum entropy method \cite{TRIQS/maxent} to analytically continue the self energy.

For the MLWF and SLWF cases, we also perform DMFT calculations using $U$ and $J$ values fitted from constrained random phase approximation (cRPA) calculations as implemented in VASP~\cite{Kaltak2015}. We use the Wannier orbitals as described above for the construction of the bare Coulomb interaction, and to evaluate the screening by splitting the polarization as $P = P_{\text{sub}} + P_{\text{rest}}$ where $P_{\text{sub}}$ is the polarizability for the correlated subspace (in our case, the Ni $3d$ orbitals) and $P_{\text{rest}}$ is for the rest of the system. From this, the screened  interaction tensor can be calculated in a local basis from the bare Coulomb interaction tensor $\hat{V}$, as $\hat{U}(\omega) = \hat{V}/[1 - \hat{V}P_{\text{rest}}(\omega)]$. Here, we limit ourselves to the static limit $\hat{U}(\omega = 0)$ of the screened interaction. For the cRPA calculation we use a $k$-point grid of $9 \times 9 \times 9$, with an energy cutoff of 500~eV, and $\sim300$ empty bands (plus 21 occupied bands). To extract symmetrized interaction parameters we average the full four index interaction tensor assuming cubic symmetry, obtaining the parameters for the Hubbard-Kanamori Hamiltonian used for the Ni-$e_g$ orbitals~\cite{vaugier2012}.

\section{Orbital content \label{sec:orb_cont}}

\begin{figure}[t]
    \centering
    \includegraphics[width = \linewidth]{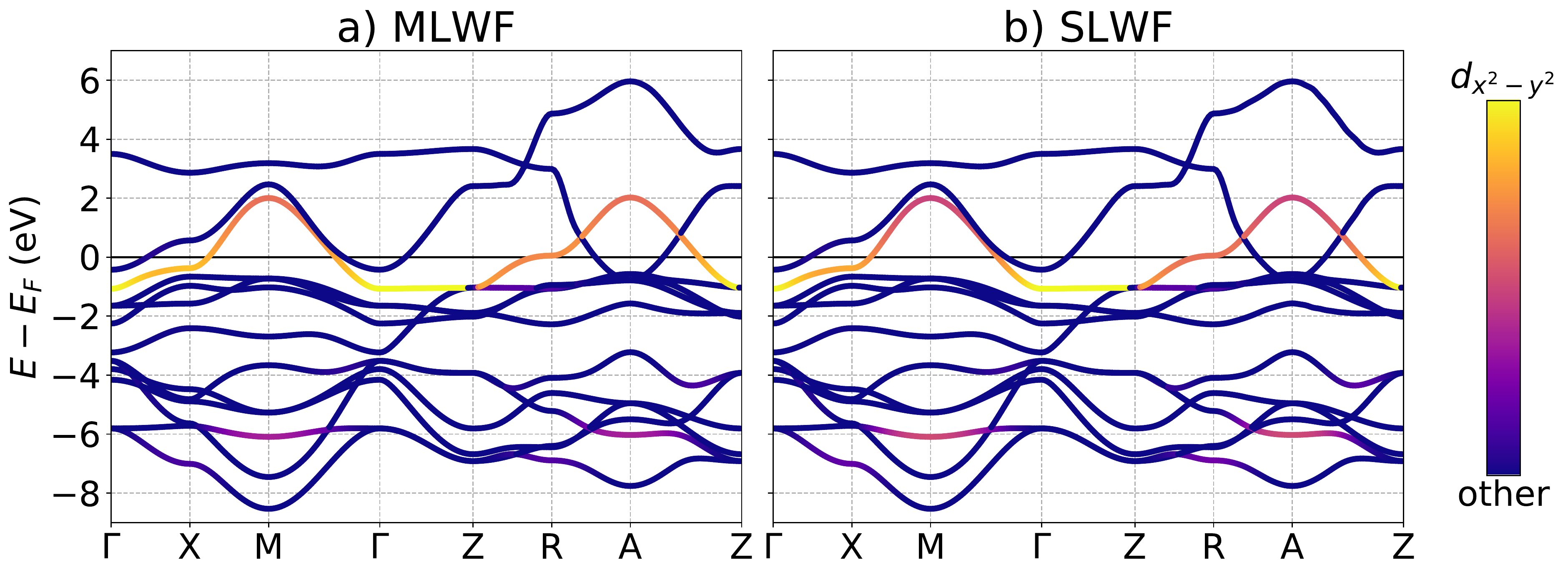}
    \caption{Energy bands obtained from diagonalizing the Wannier $H^{ab}(k)$ in the MLWF and SLWF methods. The color of the bands at each energy eigenvalue represents the amount of the $d_{x^2-y^2}$ Wannier function in the corresponding eigenvectors.}
    \label{fig:wanband_dx2y2}
\end{figure}

In this section, we examine the physical content of the Wannier and projector representations of the band theory. The Wannier methods produce an explicit representation of the Kohn-Sham bands and eigenfunctions within a given energy window and the correlated orbitals are defined as particular linear combinations of these states, permitting a straightforward analysis. Figure \ref{fig:wanband_dx2y2} shows the bands obtained from the MLWF and SLWF Hamiltonians along a high symmetry path in $k$ space, along with the $d_{x^2-y^2}$ content, indicated in pseudocolor. The energy dispersions produced in the two methods are essentially identical and are indistinguishable from the Kohn-Sham bands (not shown), but the orbital content of the bands is different in the different Wannierization schemes. The most relevant bands are the one crossing the Fermi level between $\Gamma-X-M$ and $Z-R-A$, and the weakly dispersing band at $\sim \SI{-6}{eV}$.  In the MLWF case, the Ni-$d_{x^2-y^2}$ content is more concentrated in the band that crosses the Fermi level, with less weight in the $\SI{-6}{eV}$ band, while the proportions are more equal in the SLWF method. 

\begin{table}[b]
\begin{tabular}{|l|l|l|}
\hline
 M-point $d_{x^2-y^2}$ content      &  Band at $\sim 2$ eV & Band at $\sim -6$ eV\\ \hline
MLWF   & 0.64                & 0.36                 \\ \hline
SLWF   & 0.49                & 0.51                 \\ \hline
Wien2k & 0.64                & 0.36                 \\ \hline
QE     & 0.58                & 0.42                 \\ \hline
\end{tabular}
\caption{$d_{x^2-y^2}$ content of the two bands with significant $d_{x^2-y^2}$ content at the $M$ point. In the MLWF and SLWF cases the orbital content is the modulus squared of the overlap of the band basis state with the $d_{x^2-y^2}$ Wannier function. In the Wien2k case, the orbital content is obtained from the projection of the band on the $3d_{x^2-y^2}$ basis state inside the Ni muffin tin. In the Quantum Espresso (QE) case, the projection is onto orthogonalized atomic wavefunctions. }
\label{tab:dx2y2_content}
\end{table}

Table \ref{tab:dx2y2_content} quantifies the difference, showing the Ni-$d_{x^2-y^2}$ content of relevant bands at the Brillouin zone $M$ point. We compare the orbital content obtained from the MLWF and SLWF methods  to that provided by the Quantum Espresso and Wien2k codes, which use a projector method. We see that the different methods, while exactly reproducing the energy dispersions, lead to quite different orbital contents. The difference arises because (as qualitatively seen in the two site model discussed in the previous section), the same dispersion may be fit by different tight binding parameters, which in turn lead to different orbital content of bands. Table \ref{tab:comb2-3} presents the $p$-$d$ energy difference (obtained from the orbital and site-diagonal terms of the Wannier Hamiltonian) and hybridization (the first neighbor $p$-$d$ hopping term in the Wannier Hamiltonian). We see as expected that the SLWF and MLWF methods trade off the values of $\varepsilon_d-\varepsilon_p$ and $t_{pd}$ to obtain comparable fits to the band structure.

\begin{table}[t]
\begin{tabular}{|c|c|c|c|c|c|c|c|c|}
\hline
               & $\varepsilon_d - \varepsilon_p$ & $t_{pd}$ &  & $d_{x^2-y^2}$ & $d_{z^2}$ & $d_{xz/yz}$ & $d_{xy}$ & total \\ \hline
MLWF           & 4.32                            & 1.28     &  & 1.19          & 1.83      & 1.94        & 1.97     & 8.89  \\ \hline
SLWF           & 3.38                            & 1.41     &  & 1.32          & 1.88      & 1.96        & 1.98     & 9.11  \\ \hline
Proj -10 to 10 & 2.66                            & -        &  & 1.19          & 1.58      & 1.89        & 1.95     & 8.50  \\ \hline
Proj -10 to 3  & 4.18                            & -        &  & 1.21          & 1.71      & 1.95        & 1.99     & 8.82  \\ \hline
\end{tabular}
\caption{Left: Difference between the onsite energies of the  Ni-$d_{x^2-y^2}$ and O-$p_\sigma$ Wannier functions ($\varepsilon_d - \varepsilon_p$) and hopping between them ($t_{pd}$). In the Wannier cases the parameters are read off directly from the appropiate entries in the real space Wannier Hamiltonian $H_{IJ}^{ab})$. In the projector cases, $\varepsilon_d - \varepsilon_p$ is determined by downfolding the Kohn-Sham Hamiltonian, but $t_{pd}$ is not defined. Right: Orbital occupancies of Ni-$d$ defined as the square of the projection of the occupied k-states as obtained from DFT onto the local  orbitals summed over spin.}
\label{tab:comb2-3}
\end{table}

\begin{figure}[b]
    \centering
    \includegraphics[width = \linewidth]{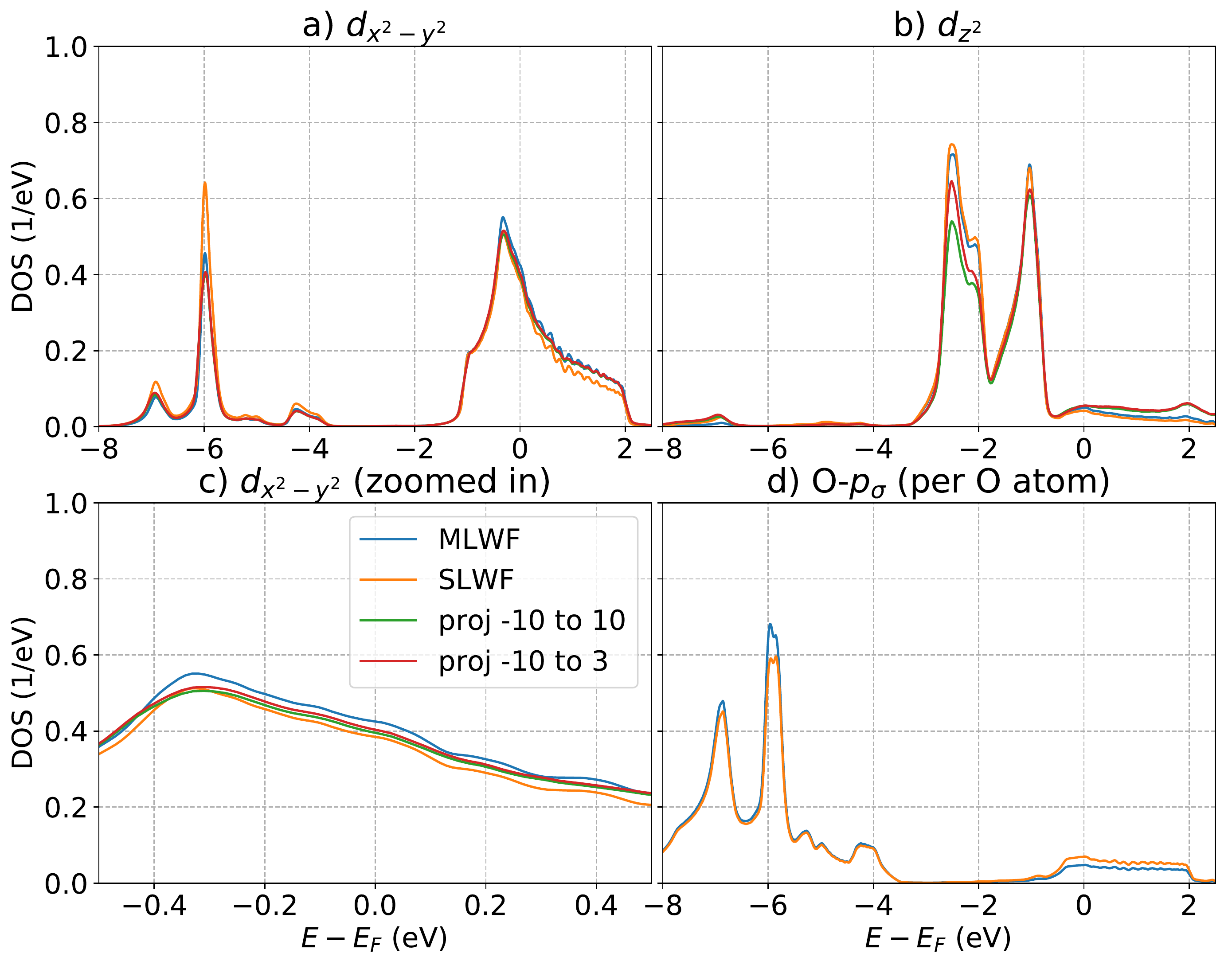}
    \caption{Uncorrelated DOS (per spin), obtained from the imaginary part of the local Green's function in the Wannier basis without self energy.}
    \label{fig:uncorrdos}
\end{figure}

Figure  \ref{fig:uncorrdos} shows the orbitally projected density of states  obtained using Wannier and projector methods. The upper left panel shows the projection  onto the $d_{x^2-y^2}$ orbital. Two peaks are observed, reflecting the strong hybridization of Ni-$d_{x^2-y^2}$ and O-$p_{\sigma}$ states which divides the $d$ density of states into bonding (low energy)  and antibonding (near Fermi energy) portions. The different methods predict different $d$ weights in the bonding (low energy) region.

Fig.~\ref{fig:uncorrdos}(b) shows the $d_{z^2}$ density of states. Around $\sim \SI{-2.5}{eV}$ a large difference between the MLWF/SLWF DOS and the projector DOS is evident, with the projector method leading to a much smaller Ni-$d_{z^2}$ density of states. Here the hybridization is not with the oxygen but with the Nd-$d$ states, as there are no oxygen states at this energy (see Fig.~\ref{fig:uncorrdos}(d)). We also see clearly that at higher energies the $d_{z^2}$ orbital is hybridized with orbitals of Nd character (not shown here) lying above the Fermi level. The different methods treat the hybridization to the higher lying states differently, and this affects the final results.

Fig.~ \ref{fig:uncorrdos} (c) shows the $d_{x^2-y^2}$ DOS in a narrow frequency range around the Fermi level. In this energy range  the MLWF $d_{x^2-y^2}$ DOS is slightly greater because it has more content from the Ni-derived band that crosses the Fermi level. However, the differences are minimal, indicating that in this case the choice of downfolding is important primarily in affecting the character of the states farther from the Fermi level.

\begin{figure}[b]
    \centering
    \includegraphics[width = \linewidth]{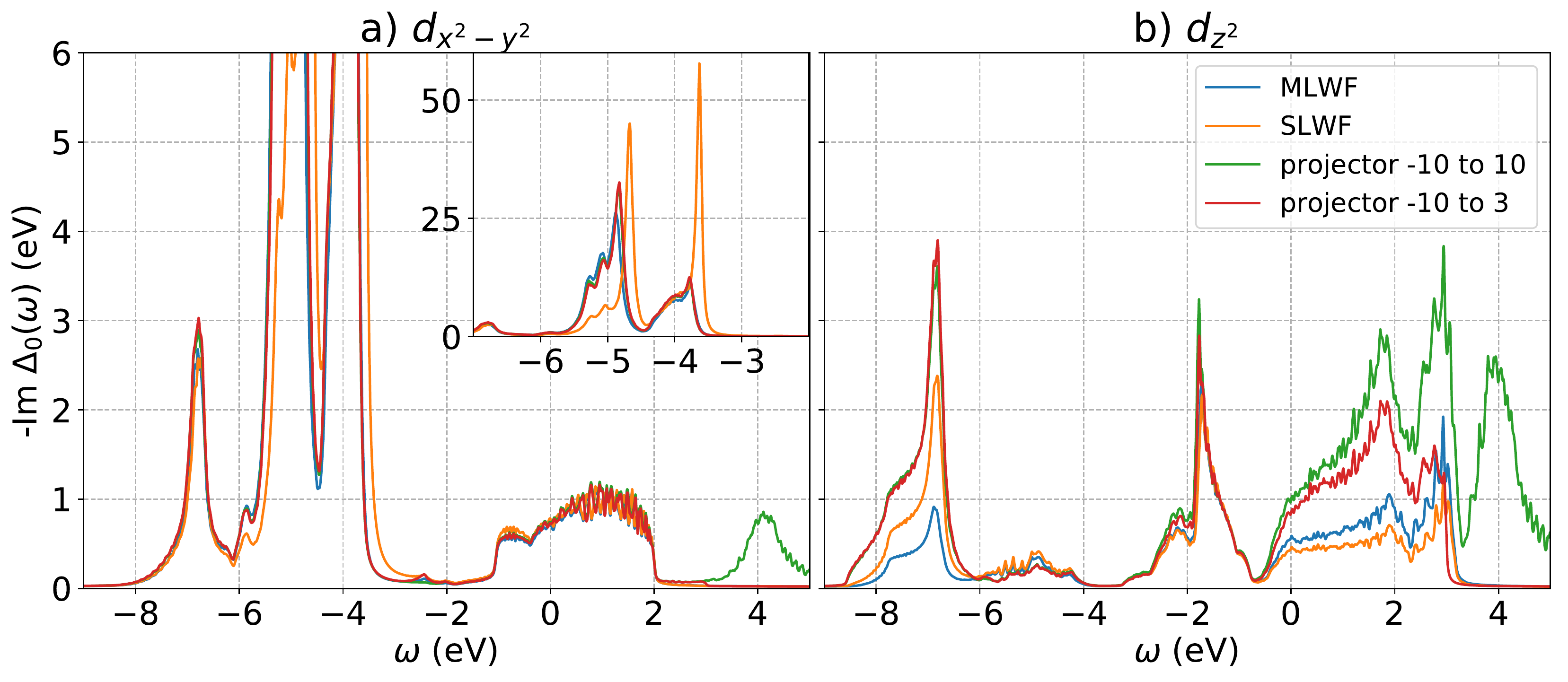}
    \caption{Negative imaginary part of the real frequency hybridization of the a) $d_{x^2-y^2}$ and b) $d_{z^2}$ orbitals. The inset of a) shows the $d_{x^2-y^2}$ hybridization zoomed in on the window of $-7$ to $\SI{-2}{eV}$ and with a larger y axis to show the large hybridization to O-$p_\sigma$. }
    \label{fig:delta0_eg}
\end{figure}

The differences are quantified in Table \ref{tab:comb2-3}, which shows the occupancy of the $d$ orbitals as obtained from each method. In all cases, the $t_{2g}$ orbitals are almost full, justifying the use of a two-orbital model that neglects correlations in these orbitals.  All methods produce a greater than half filled  $d_{x^2-y^2}$ orbital due to charge transfer from the ligand orbitals. However, the different methods lead to quantitatively different results. The $d_{x^2-y^2}$ filling is roughly the same in the MLWF and projector cases, but it is significantly greater in the SLWF case. The $d_{z^2}$ orbital is less filled in the projector than in the Wannier cases, especially the case with the larger energy window, reflecting the difference in the feature at $\sim - \SI{2.5}{eV}$ in the DOS. 

The projector methods do not provide an explicit definition of $t_{pd}$, but the physics is revealed by a comparison of the orbitally projected density of states shown in Figure  \ref{fig:uncorrdos} to the bare hybridization function, shown in Figure~\ref{fig:delta0_eg} and defined as $\hat \Delta_0(\omega) = \omega \mathds{1} - \hat \varepsilon_0 - \hat G_{loc,0}^{-1}(\omega)$, where $\hat G_{loc,0}(\omega)$ is the uncorrelated site local Green's function projected onto the basis of correlated orbitals and $\hat \varepsilon_0$ is the onsite energy obtained from $lim_{\omega\rightarrow\infty}\left(\omega \mathds{1} - \hat G_{loc,0}^{-1}(\omega)\right)$. In the simple two level model considered above, the bare hybridization function would be $t_{pd}^2/(\omega-\varepsilon_p)$. In the general case $\hat \Delta_0(\omega)$ has poles at the energies of the levels with which the correlated orbitals are hybridized, while the integrated weight ($\int Im \Delta_0(\omega)$) gives the total hybridization strength.  

Examination of the hybridization function reveals pronounced differences between the methods. In particular, Fig.~\ref{fig:delta0_eg}(a), inset shows that the SLWF method yields an intrinsically less dispersive but more strongly hybridized O-$p_\sigma$ state at a noticeably lower (less negative) energy than the other methods. For these reasons the SLWF method has a substantially larger Ni-$d$-admixture in the DOS in the  bonding energy range, reflecting the larger $t_{pd}$ and smaller $\varepsilon_d-\varepsilon_p$ found in this method. Conversely, looking at the O-$p_\sigma$ DOS (d), we see that the MLWF O-$p_\sigma$ DOS is smaller at the Fermi level but larger in the $\SI{-6}{eV}$ range of  the oxygen band energies, again reflecting that the SLWF method assigns more from the oxygen bands to the $d_{x^2-y^2}$ Wannier function. The projector method produces results in between the SLWF and MLWF methods, although closer to MLWF, indicating a smaller effective $t_{pd}$ and larger $\varepsilon_d-\varepsilon_p$ than in the SLWF method, but not quite as small (large) as in the MLWF method. 
 
\begin{table}[h]
\begin{tabular}{|c|c|c|}
\hline
               & DOS,  & $-\text{Im} \Delta_0(\omega)$ \\ \hline
MLWF           & 0.204 & 16.86                         \\ \hline
SLWF           & 0.288 & 21.41                         \\ \hline
Proj -10 to 10 & 0.213 & 17.92                         \\ \hline
Proj -10 to 3  & 0.218 & 17.98                         \\ \hline
\end{tabular}
\caption{Integral of the DOS over the energy range $-8$ to $\SI{-3}{eV}$ and hybridization function of the $d_{x^2-y^2}$ over regions of significant overlap with oxygen. }
\label{tab:integrals}
\end{table}
The differences in hybridization strength are quantified in Table ~\ref{tab:integrals}. The integral of the $d_{x^2-y^2}$ DOS over the energy region with oxygen-derived bands from $-8$ to $\SI{3}{eV}$ is greater for the SLWF case than the MLWF case, indicating a smaller effective $\varepsilon_d - \varepsilon_p$ and larger effective $t_{pd}$ in the SLWF case, in agreement with the values in table \ref{tab:comb2-3}. Likewise, the integral of $-\text{Im} \Delta_0(\omega)$ in the region of dominant peaks from $-6$ to $\SI{-3}{eV}$ is greater for the SLWF case, also indicative of a larger effective $t_{pd}$. For both of these quantities, the values in the projector cases are in between those for the MLWF and SLWF cases but closer to the MLWF case, indicating an effective $\varepsilon_d - \varepsilon_p$ slightly smaller than the MLWF case and an effective $t_{pd}$ slight larger than the MLWF case. Thus in summary we see that the selectively localized Wannier function leads to the smallest $p$-$d$ energy difference and strongest $p$-$d$ hybridization; the MLWF leads to the largest $p$-$d$ hybridization, with the projector method intermediate, but closer to the MLWF method. 

The differences in mapping from orbital to band basis in the different downfolding methods also imply a difference in interaction strengths. In the DMFT approach, the interaction parameters are chosen to represent the on-site terms in the screened Coulomb interaction. They are sometimes chosen phenomenologically or to obtain agreement with experiment (for the case of perovskite nickel oxides see \cite{Nowadnick15}),  and in most of the calculations reported in this paper these phenomenologically determined parameters are used.

\section{DMFT Results \label{sec:dmft_results}}

This section investigates the ways in which the different downfoldings lead to different results in the interacting theory. Unless otherwise specified this section presents  DMFT calculations  for a "two orbital" model of \NNO{} in which the $d_{x^2-y^2}$ and $d_{z^2}$ orbitals are considered to be dynamically correlated and the phenomenologically determined $U = \SI{7}{eV}$ and $J = \SI{0.7}{eV}$ interaction parameters discussed above are used. For comparison, we also present ``one orbital" results for the nickelates in which only the $d_{x^2-y^2}$ orbital is considered to be correlated and a brief discussion of the analogous cuprate materials, where a one band description is more widely accepted.

\subsection{Self Energy and Mass Enhancement \label{sec:dmft_sigma}}

\begin{figure}[b]
    \centering
    \includegraphics[width = \linewidth]{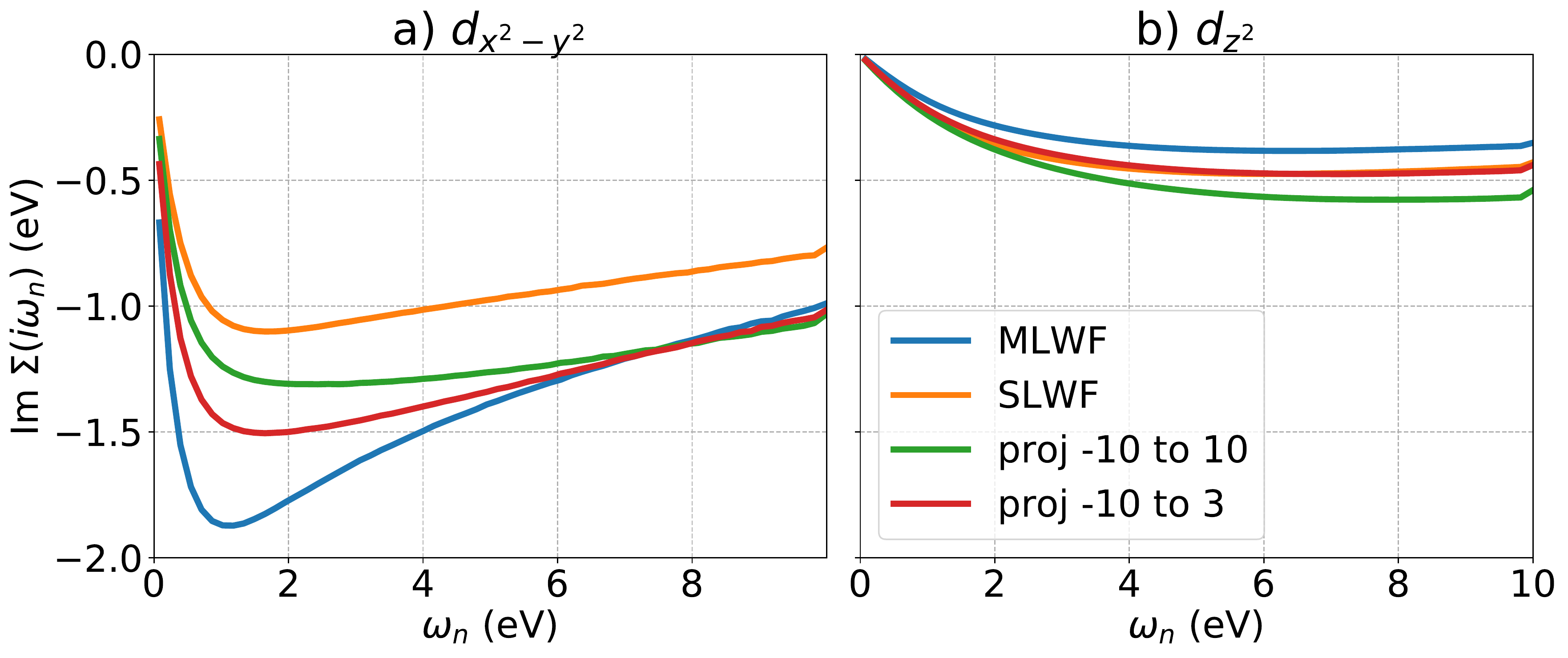}
    \caption{Imaginary part of the Matsubara self energy obtained from a DMFT solution for a two orbital Ni-$e_g$ model with a Kanamori Hamiltonian with $U = \SI{7}{eV}$ and $J = \SI{0.7}{eV}$ using the downfolding methods shown in the legends.}
    \label{fig:Sigma_iw_comp}
\end{figure}

 Figure \ref{fig:Sigma_iw_comp} compares the imaginary part of the Matsubara self energy of the different models for both the $d_{x^2-y^2}$ and $d_{z^2}$ orbitals. The difference in self energy corresponds to a difference in predicted correlation strength.  We quantify the strength of electronic correlations by the inverse quasiparticle renormalization $Z^{-1}=1-\partial Re\Sigma(\omega\rightarrow 0)/\partial \omega$ related, in the single-site DMFT approximation, to the quasiparticle mass enhancement as $m^\star/m=Z^{-1}$. At low $T$ in a Fermi liquid regime, $Z^{-1}$  can be expressed in terms of the Matsubara self energy as $Z^{-1} =  1 - \partial \text{Im}\Sigma(i \omega_n \rightarrow 0)/\partial \omega_n$. We estimate the derivative by fitting a 4th order polynomial to the first 6 Matsubara points and taking the linear term, following  ~\cite{Mravlje2011,Zingl2019}. The resulting mass enhancements are shown in Table \ref{tab:mass_and_Ueff}.
 
 \begin{table}[h]
\begin{tabular}{|c|c|c|c|c|c|}
\hline
               & $\frac{m^\star}{m}d_{x^2-y^2}$  & $\frac{m^\star}{m}d_{z^2}$ & $\frac{m^\star}{m}X$ && $U_{\text{eff}}$  $d_{x^2-y^2}$ \\ \hline
MLWF           & 7.6                  & 1.2        &6.7      &  & 3.6                             \\ \hline
SLWF           & 3.9                  & 1.3        &3.3      &  & 2.8                             \\ \hline
Proj -10 to 10 & 4.6                  & 1.3       &3.8       &  & 2.5                             \\ \hline
Proj -10 to 3  & 5.6                  & 1.3        &4.7      &  & 2.5                             \\ \hline
\end{tabular}
\caption{Left: Orbital basis mass enhancements for the $d_{x^2-y^2}$ and $d_{z^2}$ orbitals, obtained by fitting a 4th order polynomial to the first 6 Matsubara points of the imaginary part of the self energy along with  the band basis mass enhancement obtained from the quasiparticle band nearest the Fermi level at the $X$ point of the band structure.  Right: Effective $U$ values for the $d_{x^2-y^2}$, defined as the distance between the Hubbard peaks of the $d_{x^2-y^2}$ momentum integrated spectral function.}
\label{tab:mass_and_Ueff}
\end{table}

Figure \ref{fig:Sigma_iw_comp} and Table ~\ref{tab:mass_and_Ueff} show that for the $d_{x^2-y^2}$ orbital  the self energy at all frequencies as well as the mass enhancement  is much larger in the MLWF case than the SLWF case, with  projector cases being intermediate.  The $d_{z^2}$ orbital mass enhancement is  small in all cases. We attribute the differences in $d_{x^2-y^2}$ self energy and mass enhancement  to the differences in $p$-$d$ energy splitting and $p$-$d$ hybridization strength discussed in the previous section, consistent with previous literature on the charge-transfer to Mott insulator crossover \cite{Zaanen1985band,Dang14}. Note, however, that in contrast to the situations considered in previous literature, where only hybridization to oxygen bands is relevant, for \NNO{} the mass enhancements also depend on the hybridized $d_{z^2}$/Nd bands, which depend on the projection window.

\begin{figure}[t]
    \centering
    \includegraphics[width = \linewidth]{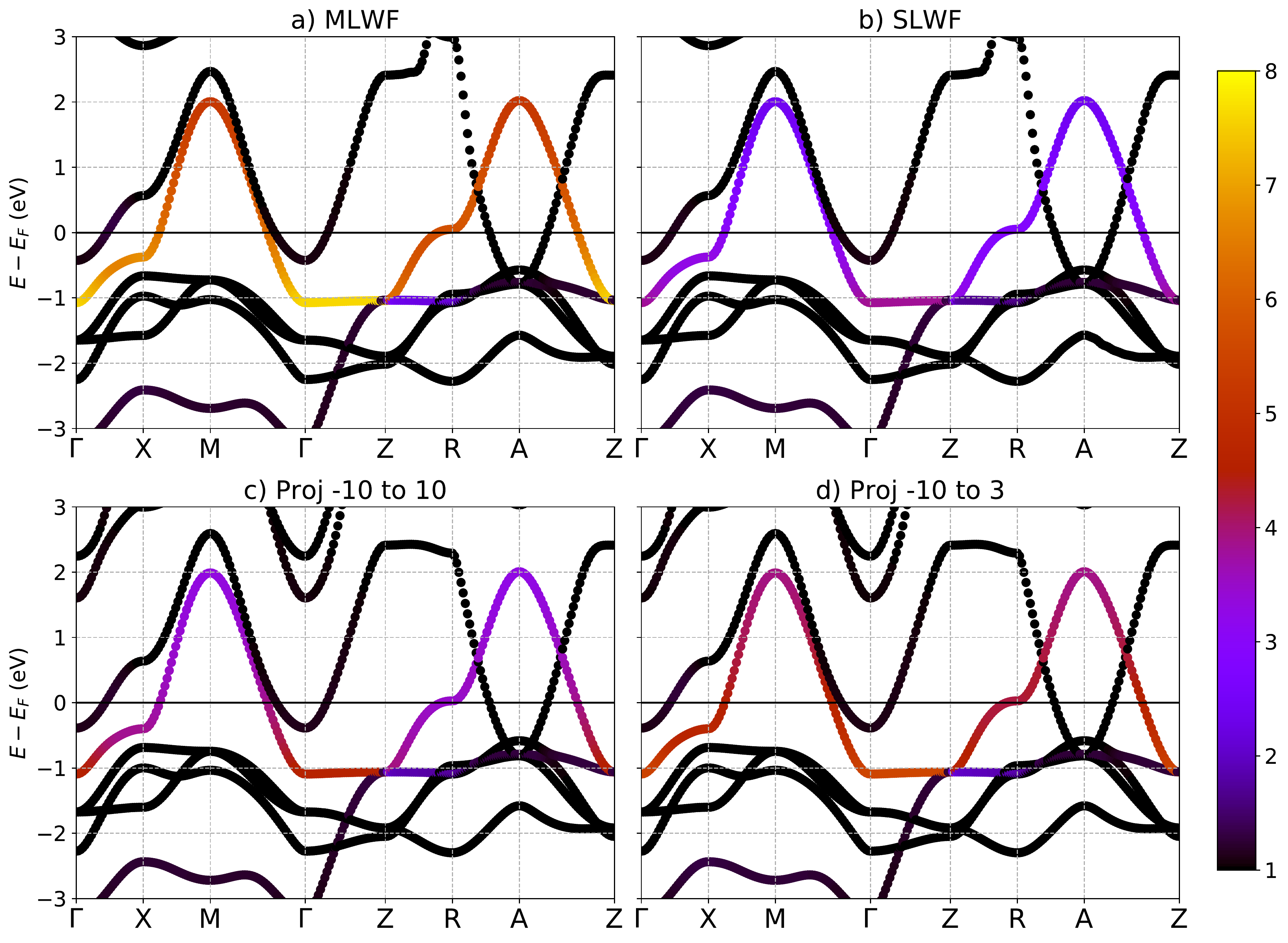}
    \caption{Pseudocolor plot of the quasiparticle mass enhancements in the band basis along a high symmetry $k$ path. In the Wannier cases the bands plotted are the same in Figure \ref{fig:wanband_dx2y2} and in the projector cases they are the  Wien2k DFT bands. The color corresponds to the mass enhancement of the DFT band determined from upfolding the self energy to the band basis. The self energy is obtained from a DMFT solution for a two orbital Ni-$e_g$ model with a Kanamori Hamiltonian with $U = \SI{7}{eV}$ and $J = \SI{0.7}{eV}$}
    \label{fig:mass_band_basis}
\end{figure}

One important caveat is that  that the values reported in Table \ref{tab:comb2-3} are  ``orbital basis'' mass enhancements, determined from the diagonal elements of the projection of the self energy operator onto the correlated orbitals. A quantity of more direct relevance to the low energy physics is the ``band basis" mass enhancement, which is proportional to the admixture of the uncorrelated orbitals in the band of interest and gives the renormalization of the quasiparticle bands with respect to the DFT bands. For the MLWF and SLWF methods, we obtain the band basis mass enhancement by transforming the self energy to the band basis using the eigenvectors of the uncorrelated Wannier Hamiltonian $H(k)$. For the projector methods, we use the projectors to upfold the self energy back to the Kohn-Sham basis (Eq. \ref{sigmaupfold}). Figure \ref{fig:mass_band_basis} shows the band basis mass enhancement along the same high symmetry path on which the bands are plotted in Figure \ref{fig:wanband_dx2y2}, and Table~\ref{tab:comb2-3} gives the value of the band basis mass enhancement for the near Fermi surface state at the X point. In the band basis, the difference between the MLWF and SLWF cases is even greater than in the orbital basis, for the same reason-- in the MLWF case there is less admixture of oxygen in the near Fermi surface band so the $d$ self energy has a greater effect on the dispersion. 

Another potential caveat is that different choices for downfolding may lead to different interaction parameters. To investigate the basis dependence of the interaction parameters we have  used the  ``constrained Random Phase Approximation" (cRPA) approach to estimate the Coulomb parameters corresponding to the two Wannier downfoldings. This approximation is believed to underestimate the true interactions, but  gives trends correctly. Symmetrizing our compouted screened Coulomb tensor  over the two active  orbitals gives parameters  $U \approx \SI{3.6}{eV}$ and $J \approx \SI{0.7}{eV}$ for MLWF and $U \approx \SI{2.9}{eV}$ and $J \approx \SI{0.7}{eV}$ for SLWF. We observe  that although the SLWF correlated orbitals are smaller (less spatial extent) than the MLWF orbitals,  the SLWF approach yields smaller interaction parameters than does the MLWF approach, because the dominant effect on the interaction parameters is from screening, which is stronger for the smaller $\varepsilon_d-\varepsilon_p$ found in the SLWF method. Use of the  cRPA interaction parameters (smaller for SLWF and for MLWF)  yields a Ni-$d_{x^2-y^2}$ orbital mass enhancement of $2.1$ for MLWF and $1.4$ for SLWF, corresponding to a factor of almost $3$ in correlation contributions $m^\star/m-1$. Thus the difference in interaction parameters arsising from differences in downfold amplify, rather than decrease, the downfolding-induced  differences in self energy.

\subsection{Spectral function \label{sec:dmft_spectral}}
\begin{figure}[h]
    \centering
    \includegraphics[width = \linewidth]{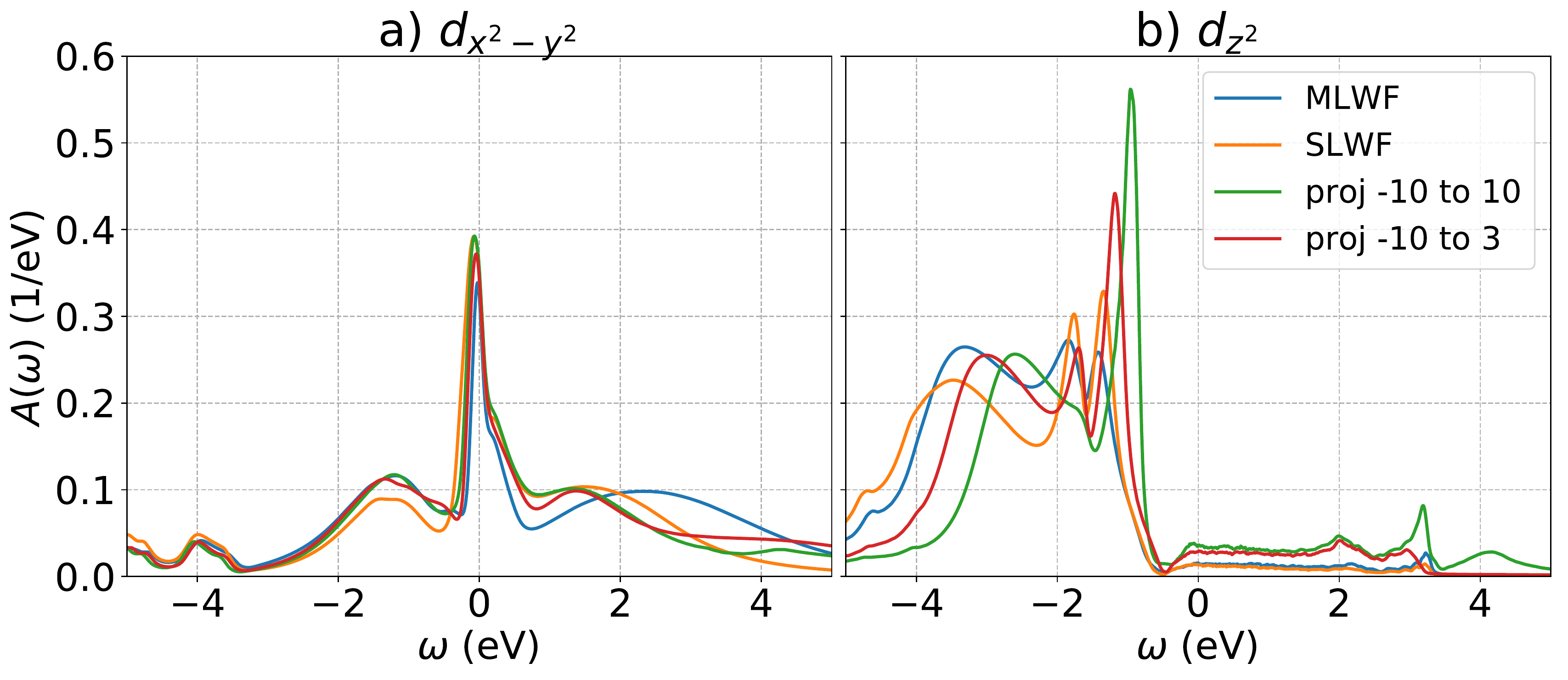}
    \caption{Momentum integrated spectral functions per spin obtained from DMFT solutions for a two orbital Ni-$e_g$ model with a Kanamori Hamiltonian with $U = \SI{7}{eV}$ and $J = \SI{0.7}{eV}$ using the downfolding methods shown in the legends.}
    \label{fig:Aw}
\end{figure}

Figure \ref{fig:Aw} shows the orbitally resolved momentum integrated spectral function ${\hat A}(\omega) = i\left[{\hat G}(\omega) - {\hat G}(\omega)^\dag\right]/2\pi$ for the $d_{x^2-y^2}$ and $d_{z^2}$ orbitals  for a range of energies not too far from the chemical potential.  The  $d_{x^2-y^2}$ orbital (left panel) exhibits a three peak structure similar to that found in the single-band  Hubbard model at moderate correlation strength. Interpreting the structure in terms of a low energy effective model, we identify the electron removal peak at $\omega\approx \SI{-1.5}{eV}$ with the lower Hubbard band, the broader peak at $\sim \SI{2}{eV}$ with the upper Hubbard band, and the central peak near $\omega=0$ with the quasiparticle band. The energy separation between the lower and upper ``Hubbard peaks", shown in Table \ref{tab:mass_and_Ueff} then provides an estimate for the effective interaction $U_{eff}$  characterizing an effective low energy model. While all methods provide qualitatively similar spectral functions,  in the MLWF case the Hubbard peaks are further away from each other, indicating a greater effective Hubbard repulsion due to the greater $\varepsilon_d-\varepsilon_p$ and smaller $t_{pd}$  and consistent with the larger mass enhancement found in the previous subsection. 

The spectral function for the  $d_{z^2}$ orbital  shows a weak tail at energies above the chemical potential, a sharp peak at $\sim \SI{-1}{eV}$ and a broad feature in the range $\sim -2$ to $\sim \SI{-4}{eV}$. Again all methods produce the same qualitative behavior, but differ quantitatively.  In the projector cases the peaks are closer to the chemical potential and there is more weight above the Fermi level than in the Wannier cases. 

\begin{figure}[h]
    \centering
    \includegraphics[width = \linewidth]{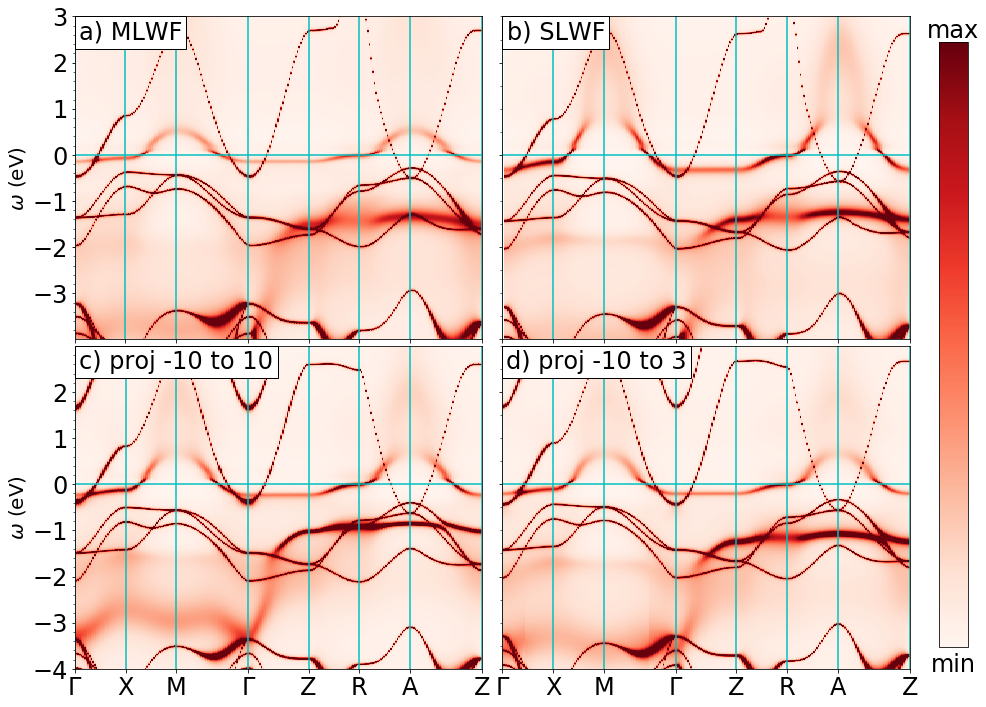}
    \caption{Pseudocolor plots of the momentum resolved spectral functions $A(k,\omega)$ obtained from DMFT solutions for a two orbital Ni-$e_g$ model with a Kanamori Hamiltonian with $U = \SI{7}{eV}$ and $J = \SI{0.7}{eV}$.}
    \label{fig:Akw}
\end{figure}

Figure \ref{fig:Akw} shows the momentum resolved spectral function $A(k,\omega)= -\text{Tr}\left[\text{Im }\hat{G}(k,\omega)\right]/\pi$ along a high symmetry path in the Brillouin zone. The bands with significant correlation effects appear more diffuse because of the larger imaginary part of the self energy. Comparison to Fig ~\ref{fig:Aw} shows that the correlated band crossing the Fermi level (e.g. between X and M) arises from the $d_{x^2-y^2}$ orbital.  Corresponding to the different mass enhancements, this  band  is renormalized the most in the MLWF case (see e.g. the distance below the Fermi level of the band at the Z point), then the projector cases, and then the SLWF case. The sharp peak visible in the  $d_{z^2}$ density of states in Fig. ~\ref{fig:Aw} arises from the almost dispersionless correlated band visible from Z to A, while the features in the $-\SI{2}{eV}$ to $\SI{-4}{eV}$ range arise from the more diffuse features seen between $\Gamma$ and M in the momentum-resolved figures. The energy position and relative sharpness of these features depends on the downfolding method, providing the possibility of experimental tests of different downfoldings.

\subsection{Orbital Occupancies \label{sec:dmft_orb_occ}}

The occupancy of the different orbitals has been viewed as an important diagnostic of correlation physics. For example, in a model with a single relevant orbital, the orbital is more Mott-Hubbard like as it gets closer to half filling, while an occupancy noticeably greater than half filling implies important charge transfer effects. Conversely, a significant probability of occupation (by holes) of more than one orbital is a necessary condition for Hund's metal physics.  Here we consider how the calculated orbital occupancies depend on the downfolding methodology.

\begin{table}[t]
\begin{tabular}{|c|c|c|c|c|c|c|c|}
\hline
               & $d_{x^2-y^2}$ & $d_{z^2}$ && LS N=2 & HS N=2 & N=3  & N=4  \\ \hline
MLWF           & 1.13          & 1.91      && 0.04   & 0.05   & 0.78 & 0.13 \\ \hline
SLWF           & 1.27          & 1.93      && 0.03   & 0.02   & 0.69 & 0.26 \\ \hline
Proj -10 to 10 & 1.14          & 1.65      && 0.11   & 0.15   & 0.64 & 0.09 \\ \hline
Proj -10 to 3  & 1.15          & 1.81      && 0.07   & 0.08   & 0.72 & 0.12 \\ \hline
\end{tabular}

\caption{Left: Orbital occupancies of the correlated orbitals, obtained from the Matsubara Green's function. Right: Occurrence probabilities of multiplet configurations obtained from the impurity density matrix. LS stands for low spin ($s = 0$) and HS stands for high spin ($s = 1$).}
\label{tab:dmft_occ}
\end{table}

The left side of Table \ref{tab:dmft_occ} shows the  orbital occupancies, obtained directly from the impurity Green's function $\hat G_{QI}(i\omega_n$) without analytic continuation. Comparison to Table \ref{tab:comb2-3} shows that in all methods the main effect of adding correlations is to drive  the $d_{x^2-y^2}$ orbital  closer to half filling while  the $d_{z^2}$ orbital gets more full. However, the $d_{z^2}$ orbital is  significantly less full in the projector cases than in the Wannier cases, with the filling depending on the energy window employed and being smallest  for the wider window extending to $\SI{10}{eV}$. This demonstrates the importance of the hybridization to the Nd orbitals at positive energy.

The right side of Table \ref{tab:dmft_occ} shows the probabilities of different multiplet configurations of the correlated states obtained from the impurity density matrices determined from the CTHYB solver. Important quantitative differences are evident. In all cases the $N=3$ configuration is dominant, but in the Wannier cases the fluctuation into $N=4$ (fully occupied $e_g$, spin singlet) are larger than the fluctuations into $N=2$ (2 holes in the $e_g$; potential high spin state), whereas in the projector methods the situation is opposite. These differences have been used to argue for and against the relative importance of Hund's and charge-transfer physics. \cite{karp2020manybody, karp2020comparative, wang2020hunds, kang2020optical, petocchi2020normal, liu2020doping}

\subsection{One vs. Two Orbital Results \label{sec:dmft_one_orb}}

One way to approach the physics of a complicated material such as the multilayer nickelates is to attempt to define a ``minimal  model" of the  correlation effects. The much larger value of the $d_{x^2-y^2}$ self energy than the $d_{z^2}$ self energy suggests that a minimal model might involve only one correlated orbital. Insight into this possibility may be obtained by comparing results obtained from a model with multiple correlated orbitals to those obtained from a model with only one correlated orbital.

\begin{figure}[h]
    \centering
    \includegraphics[width = .8 \linewidth]{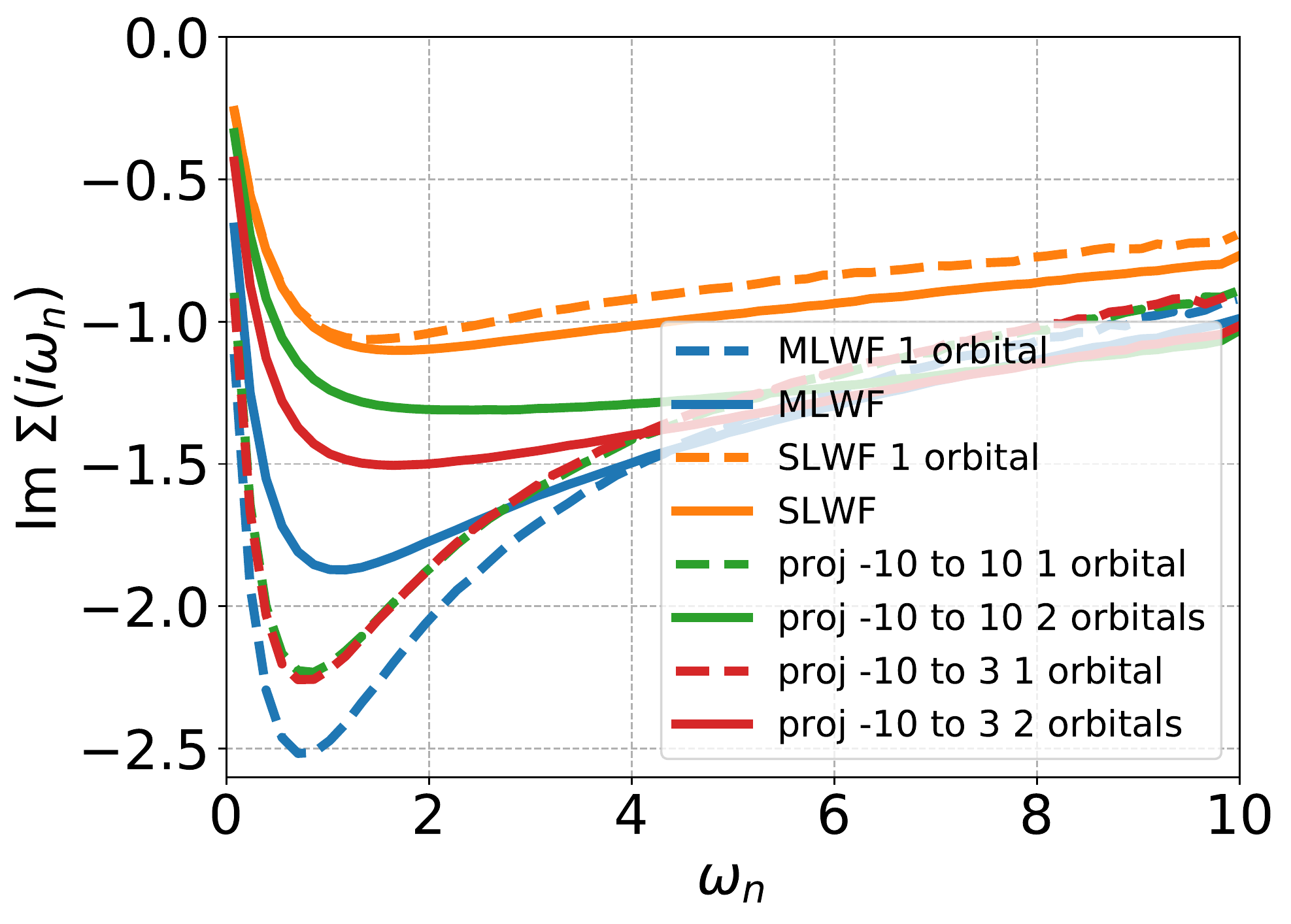}
    \caption{Comparison of imaginary part of the Matsubara self energy of the $d_{x^2-y^2}$ orbital obtained from two orbital DMFT calculations (solid lines) to self energy obtained from one orbital DMFT calculation (dashed lines) with downfolding methods unchanged.  }
    \label{fig:1v2}
\end{figure}

We use exactly the same downfolding, definition of correlated orbitals, and interaction $U$ as in our previous two orbital DMFT calculations to perform ``one orbital" DMFT calculations in which  only the $d_{x^2-y^2}$ orbital is treated as correlated. Figure ~\ref{fig:1v2} compares the resulting $d_{x^2-y^2}$ self energies, and Table~\ref{tab:1vs2mass} compares the mass enhancements. In the SLWF case, the $d_{x^2-y^2}$ self energy and mass enhancement are not changed considerably by including the $d_{z^2}$ orbital, presumably because the $d_{z^2}$ orbital is almost completely full. In the MLWF and projector cases, there is a significant difference in the self energies and mass enhancements between the one and two orbital results. Referring to Tables ~\ref{tab:dmft_occ} and  ~\ref{tab:1vs2mass} we attribute the differences to a combination of the difference from half filling of the $d_{x^2-y^2}$ orbital (with the larger occupancy in the SLWF case indicating a greater relevance of charge transfer physics) and (especially in the projector cases) greater number of holes in the $d_{z^2}$ orbital.  The SLWF results imply  that correlation physics related to the $d_{z^2}$ orbital may be neglected without adversely affecting the accuracy of the results; the other methods would suggest that this is not the case.

\begin{table}[b]
\begin{tabular}{|l|l|l|l|l|}
\hline
               & $\frac{m^\star}{m}$ 1 orb & $\frac{m^\star}{m}$ 2 orb& $n_{x^2-y^2}$ 1 orb& $n_{x^2-y^2}$ 2 orb\\ \hline
MLWF           & 11.0      & 7.6    &1.14 &1.19    \\ \hline
SLWF           & 3.8       & 3.9    &1.29 &1.32    \\ \hline
proj -10 to 10 & 9.7       & 4.6     &1.14 &1.19   \\ \hline
proj -10 to 3 & 9.7       & 5.6   &1.16 &1.21     \\ \hline
\end{tabular}
\caption{Comparison of the mass enhancement and filling of $d_{x^2-y^2}$ orbital in the one and two orbital DMFT calculations on \NNO{}.}
\label{tab:1vs2mass}
\end{table}

\subsection{Cuprate results \label{sec:dmft_cuprate}}

The layered $d^9$ nickelates of primary interest in this paper have a potentially rich physics associated with multiple bands at the Fermi surface and important ligand states lying both above and below the strongly correlated $d$ states. In this subsection we examine the extent to which our qualitative considerations apply also to the electronically simpler cuprate system, where only the $d_{x^2-y^2}$ is correlated (the $d_{z^2}$ orbital is to good approximation completely full) but charge transfer to oxygen is also relevant. In this examination, a difficulty immediately arises. A straightforward application of the DFT+DMFT methodology outlined above predicts a rather weakly correlated system, essentially because $\varepsilon_d-\varepsilon_p$ is so small in magnitude that the $d$-$d$ interaction $U$ is irrelevant. Previous work has argued that straightforward application of the DFT+DMFT method is not appropriate, essentially because the DFT approximation predicts that the oxygen levels are about $\SI{1}{eV}$ closer to the Fermi level than they are in practice.  Adjusting the $p$-level energy ``by hand" to match photoemission experiments \cite{Wang12} provides cuprate correlation physics in better agreement with experiment and we follow this route here.

\begin{table}[]
\begin{tabular}{|c|c|c|c|c|c|c|}
\hline
     & $\frac{m^\star}{m}$ $d_{x^2-y^2}$ &  & $n$ $d_{x^2-y^2}$ &  & N=3  & N=4  \\ \hline
MLWF & 1.9                               &  & 1.41              &  & 0.56 & 0.42 \\ \hline
SLWF & 1.5                               &  & 1.52              &  & 0.45 & 0.54 \\ \hline
\end{tabular}
\caption{Results for the $d_{x^2-y^2}$ orbital from a two orbital calculation on \CCO{} where $\varepsilon_p$ is reduced by \SI{1}{eV}. Left: Mass enhancement Middle: Orbital occupancy. Right: Occurrence probabilities of multiplet configurations.}
\label{tab:cuprate_epdown_dx2y2}
\end{table}

Table ~\ref{tab:cuprate_epdown_dx2y2} shows that the phenomenon found in theories of the nickelate materials occurs also in theories of cuprates: the MLWF method yields substantially larger $d_{x^2-y^2}$ mass enhancements than does the SLWF method. Examination of the Wannier fits (not shown) reveals that origin is the same--the MLWF parametrization corresponds to a larger $\varepsilon_d-\varepsilon_p$ and smaller $t_{pd}$ and hence to stronger correlations. 

The occupancy analysis shown in Table  ~\ref{tab:cuprate_epdown_dx2y2} confirms this conclusion, revealing a larger covalence (more $N=4$ weight) for SLWF than for MLWF.  It is important to note, however, that in contrast to the nickelate case, where the different methods point towards different underlying physics, in the cuprate case both the MLWF and the SLWF methods paint the same picture of a charge transfer material, different only in quantitative aspects.

\section{Discussion \label{sec:discussion}}

Quantum embedding methods approach the correlated electron problem by defining a subset of ``correlated orbitals" whose contributions to the physics are determined by the use of a high level many body method and are self-consistently embedded into a more complex electronic structure specified by an inexpensive, lower-level method. In the DFT+DMFT approach the correlated orbitals are identified as partly filled, atomic-like orbitals relatively tightly localized to particular ions. Implementation of this appealing idea encounters the difficulty that the intuitively clear idea of a ``transition metal $d$ orbital" cannot be defined unambiguously because an atomic-like state is not an eigenstate of any reasonable single particles approximation to the electronic Hamiltonian. What must be done, in effect, is to define a single particle basis that includes atomic-like states with the desired spatial structure and enough other states so that the projection of the Kohn-Sham Hamiltonian onto this basis reproduces the DFT band structure. Different methods have been used to define the correlated orbitals and select the additional states; see  Sec.~\ref{Methods} for a discussion of the principal techniques. While the issue seems not to have been extensively investigated (see \cite{park2014computing} for an exception), the consensus in the field has been that all methods that produce correlated states with approximately the desired spatial structure and reproduce the DFT bands accurately are approximately equally good.  This paper investigates the issue carefully and finds that this is not at all the case, because  the different methods in effect  lead to different  partitioning of the band states into  correlated and uncorrelated components, and these differences in partitioning have a substantial effect on the computed correlation physics.  

Focusing on one system of intense current interest, the layered  nickelate \NNO{}, this paper performs a comparative study of the implications for the many body physics of the methodology used to construct the correlated orbitals. In the layered nickelates the important correlation physics is believed to relate to states arising from the Ni $d_{x^2-y^2}$ and perhaps also from $d_{z^2}$ orbitals. Straightforward quantum chemical considerations suggest that the primary valence configuration of the Ni is $d^9$ with one hole in the $d_{x^2-y^2}$ orbital, but other configurations may also be important.  Currently debated questions include the relevance of ``Hund's metal" physics arising from high-spin $d^8$ (two holes, one in $d_{x^2-y^2}$ and one in $d_{z^2}$), the importance of correlation physics in the Ni/Nd hybrid bands crossing the Fermi level, and the relevance of valence fluctuations involving the O-${p}$.  In addressing these questions, the definition of the Ni-$d$ states and their hybridization to other orbitals is evidently crucial.

We use different variants of the two most widely used techniques, the projector and Wannier methods discussed in detail in Sec.~\ref{Methods}, to compute various physical quantities, while keeping everything else the same.   In Sec.~\ref{sec:dmft_sigma} we show that different methods lead to almost factor-of-two differences in the predicted renormalization factor (``mass enhancement"). In Sec. ~\ref{sec:dmft_orb_occ} we show that the different methods also give quite different predictions for the relevance of multiorbital (high spin, ``Hund's metal") physics, and this conclusion is reinforced in Sec. ~\ref{sec:dmft_one_orb} which shows that the different methods give also very different results for the changes in many body properties between models with two and one correlated orbitals. The issues are not specific to the \NNO{}: Sec. ~\ref{sec:dmft_cuprate} shows that similar results are obtained in a model of the copper-oxide superconductor \CCO{}, where only one correlated orbital is relevant and charge transfer physics plays a larger role.

Before proceeding to the discussion of origin and implications of the results, we dispose of two side issues. First, the results mentioned in the previous paragraph all pertain to properties of the correlated orbitals as defined in the different methodologies. The  correlated orbitals themselves are only  auxiliary quantities used in intermediate stages of computations of experimental observables. Physically meaningful results are experimental observables such as  the mass enhancement, relative to the underlying DFT mass, of the theoretically derived  quasiparticle bands (see Fig.~\ref{fig:Akw}), which are measurable in angle-resolved photoemission.  Table ~\ref{tab:mass_and_Ueff} shows that the differences between these  ``band basis'' mass enhancements are actually greater than the orbital basis self energy renormalizations. One may similarly consider the many-body density of states (local spectral function) measurable in angle-integrated photoemission and analyzed in Sec. ~\ref{sec:dmft_spectral}. We find that  large differences between methods also appear in the local  spectral function. Of special interest are the upper Hubbard feature around $\SI{2}{eV}$ and the  strong shift of the as Ni-$d_{z^2}$ characterized band between the different methods (see Fig.~\ref{fig:Aw}).

Second, most of our calculations investigate variations at fixed values of the interaction parameters, while different specifications of the correlated orbitals also imply differences in the interaction parameters governing the physics of these orbitals, which might compensate to some degree for the differences in orbital specification.  We examine this issue in Sec. ~\ref{sec:dmft_sigma}, which presents results of cRPA calculations of effective interaction parameters for different downfolding schemes. We find that the differences in interaction parameters arising from differences in specification of correlated orbitals are such as to enhance the differences between methods. In summary, different prescriptions for defining correlated orbitals lead to different results for physically measurable quantities.

We now discuss the interpretation and implications of our results. Sec.~\ref{sec:orb_cont} shows that the different parameterizations lead to quite different fillings of the correlated orbitals, already on the DFT level. Orbital filling is an important determinant of correlation physics; for example, Hund's metal physics requires at least two partially filled orbitals while Mott physics is most  pronounced if there is one nearly half filled correlated orbital. We find that the MLWF Ni-$d_{x^2-y^2}$ occupation is much closer to half-filling than the SLWF Ni-$d_{x^2-y^2}$ occupation, and is thus more likely to be found more correlated. Both projector approaches have very similar Ni-$d_{x^2-y^2}$ occupation compared to MLWF. However, the Ni-$d_{z^2}$ content is lower for projectors, depending on the energy window, which could be an indicator that projectors fail to capture interstitial contributions in strongly hybridized systems.  This is observed both with Wien2k and VASP projectors. 

The differences arise because different constructions of the correlated orbitals correspond to different embeddings of the correlated orbital in the underlying band theory, or in other words to different overlaps of correlated  orbitals with Kohn-Sham eigenfunctions. To be explicit, for transition metal oxides such as \NNO{}, important parameters include  the energy level difference between oxygen $p$ and transition metal $d$ orbitals $\varepsilon_d-\varepsilon_p$, and the $p$-$d$ hybridization $t_{pd}$ (see Tab.~\ref{tab:comb2-3}) and (see Tab.~\ref{tab:integrals}). Different constructions of the correlated orbitals lead to equally accurate parametrizations of the calculated band structures but with drastically different values of $\varepsilon_d-\varepsilon_p$ and $t_{pd}$. Furthermore, we find that the hybridization to the Nd-$d$ orbitals right above the Fermi level is strongly method dependent  (see Fig.~\ref{fig:delta0_eg}). We emphasize that this behavior is found both for VASP and Wien2K band theory codes. 

Section ~\ref{sec:dmft_orb_occ} shows that the different methods lead to differences in the multiplet occurrence probabilities in the interacting theory. The multiplet occurrence probabilities are often used to gauge the nature of correlations. In a typical one orbital Mott-Hubbard system we expect a dominance of $N = 3$ with roughly equal amounts of $N = 4$ and low spin $N = 2$. In a charge transfer material, we expect more $N = 4$ than $N = 2$. This is seen in the MLWF case and to a much greater extent in the SLWF case, but not in the projector cases. Conversely, a large amount of high spin $N = 2$, seen in the projector cases but not the Wannier cases, points to an importance of Hund's correlations. Using these results to classify the material will therefore lead to different conclusions based on the downfolding method employed, and may explain the differences in classification found in the literature. Likewise, our results in Section \ref{sec:dmft_one_orb} show that comparing one and two orbital calculations using the different methods leads to different results on the importance of multiorbital effects. 

The results presented in this paper, along with the existing discussions in the literature around the issues of value and frequency dependence of interaction parameters and the double counting correction, underscore the fact that the DFT+DMFT methodology requires choices at various points in the calculation. As noted in other contexts \cite{Nowadnick15}, experiment can to some degree help to guide the required choices. Experimental probes involving form factors that can distinguish between $d$ and $p$ orbitals can help pin down orbital content of different bands. Further, differences in  e.g. the position of the $d_{z^2}$ orbital relative to the chemical potential (see Fig. ~\ref{fig:Aw}) and broadening of the $\SI{-3}{eV}$ band (Fig.~\ref{fig:Akw}) also distinguish the methods. 

On the theoretical side, our work highlights the importance of the development of  methods \cite{Lan16,Zhu20} which treat more of the orbitals as correlated and include more of the matrix elements of the Coulomb interaction. While not all of these calculations are yet in a position to treat the strong correlation problem, comparison even in a more weakly correlated limit will provide insight.  Further, as a simple benchmark, in new situations the robustness of the results with respect to different downfolding methods should be verified. 

It is important to emphasize that the differences we highlight are in many cases quantitative rather than qualitative. For example, all methods place the cuprate materials firmly in the class of charge transfer compounds, whereas the nickelate materials are closer to the Mott-Hubbard regime. Further, the results presented here are important only for ``wide window" calculations involving both correlated and uncorrelated orbitals, and are more significant for materials such as the layered nickelates that exhibit a rich interplay between strongly and weakly correlated orbitals and between single band and multiband effect.  In cases such as Sr/Ca$_2$RuO$_4$ where the correlated bands are to good approximation disentangled from the other bands and a low energy theory involving only the correlated bands may be constructed, the physics is independent of the method used to construct the correlated orbitals. 

The success of DFT+DMFT in many contexts motivates further research to determine the optimal downfolding approach for different physical contexts. Identification of experimental observables that will distinguish different downfoldings will also be valuable, as would be the determination of quantities that are robust with respect to choice of downfolding. 

\section{Acknowledgements}
J.K. and A.J.M. acknowledge funding from the Materials Sciences and Engineering Division, Basic Energy Sciences, Office of Science, US DOE. We thank F. Lechermann, S. Beck, M. Zingl, A. Botana, and A. Georges for very helpful discussions.  The Flatiron Institute is a division of the Simons Foundation.

\appendix

\section{$t_{2g}$ DOS and Hybridization}
\label{app:t2g}

\begin{figure}[h]
    \centering
    \includegraphics[width = \linewidth]{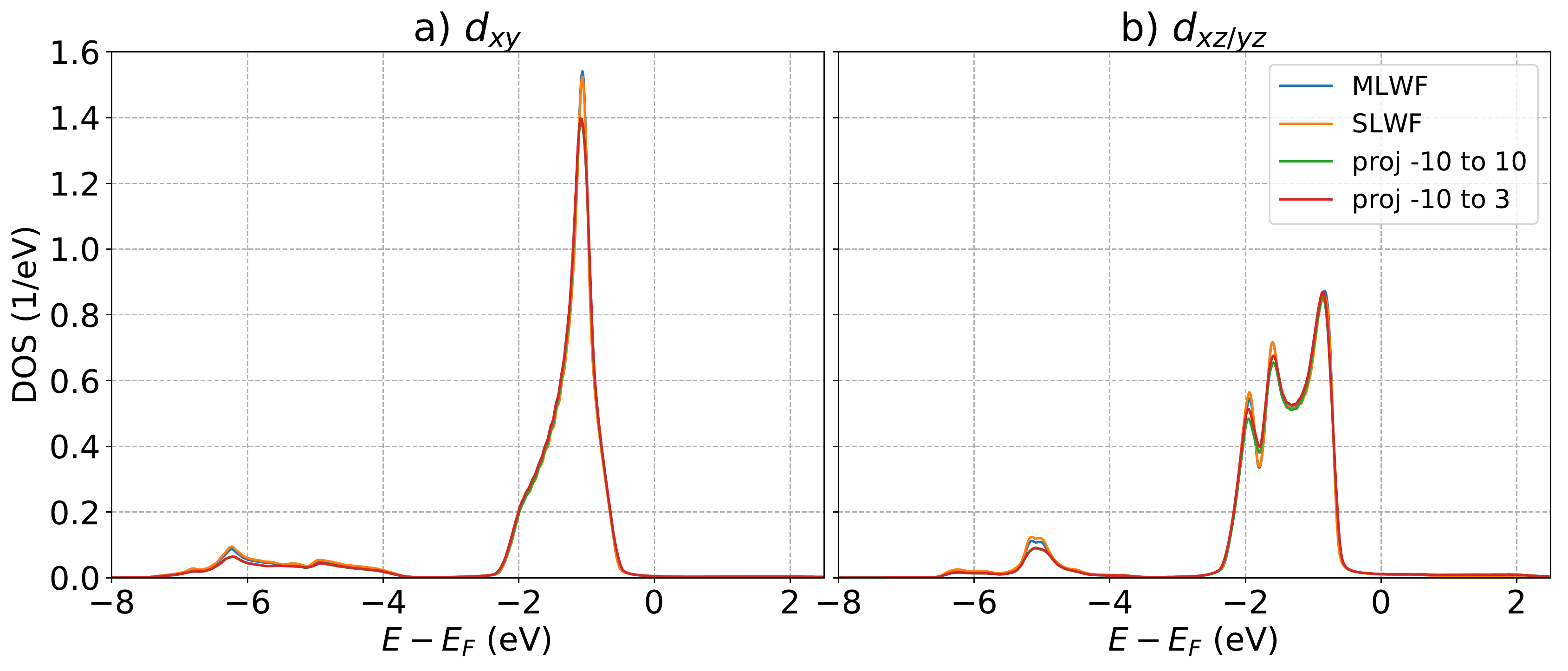}
    \caption{Uncorrelated density of states (per spin) of the $t_{2g}$ orbitals with the different methods.}
    \label{fig:uncorrdos_t2g}
\end{figure}

\begin{figure}[h]
    \centering
    \includegraphics[width = \linewidth]{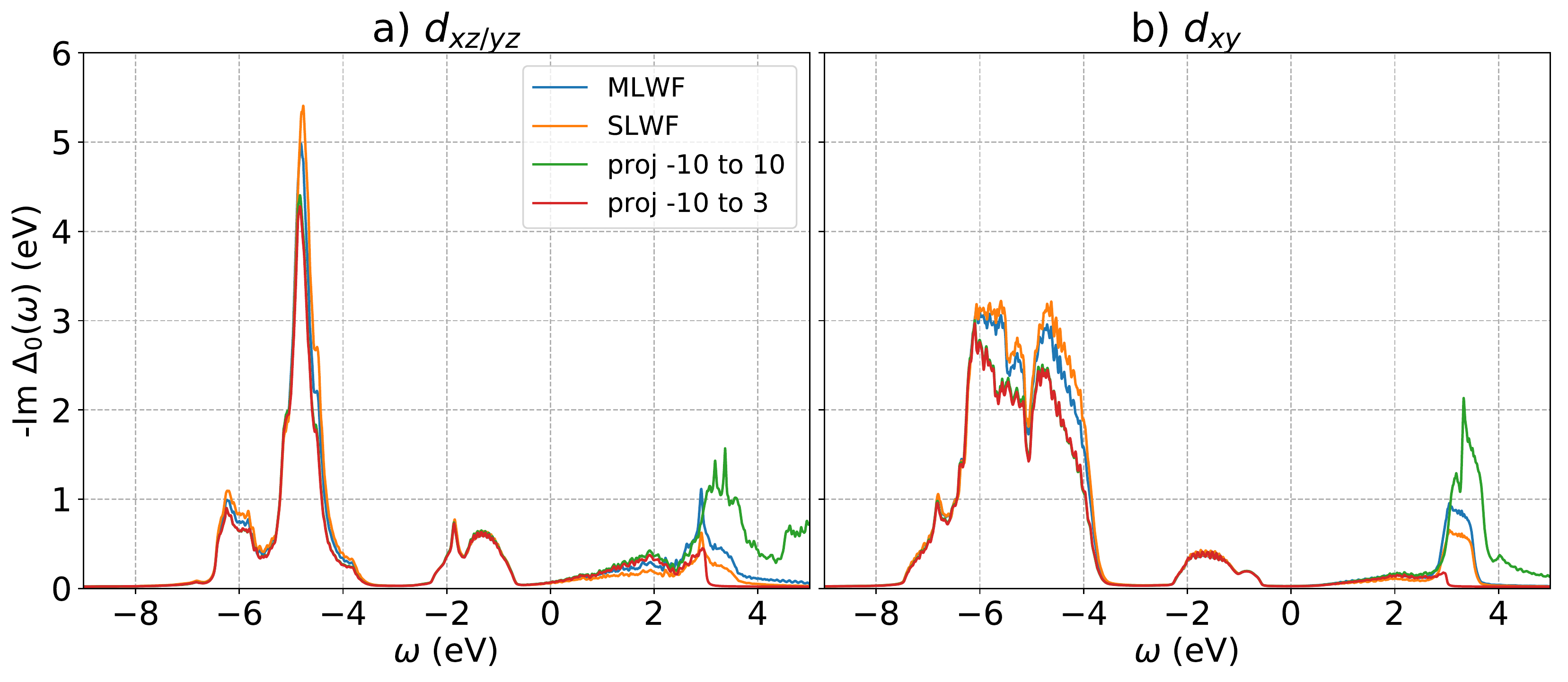}
    \caption{Negative imaginary part of the real frequency hybridization of the $t_{2g}$ orbitals.}
    \label{fig:delta_t2g}
\end{figure}

Figure \ref{fig:uncorrdos_t2g} show the density of states and Figure \ref{fig:delta_t2g} show the imaginary part of the real frequency hybridization for the Ni-$t_{2g}$ orbitals, corresponding to the plots for the $e_g$ orbitals in Figures \ref{fig:uncorrdos} and \ref{fig:delta0_eg}. The plots show some differences between methods, but less drastic than the $e_g$ orbitals. 

\section{Doping Dependence}

To simulate the effects of hole doping, we use Wien2k with the virtual crystal approximation, adjusting the atomic numbers of the Nd ions to fractional values and correspondingly change the number of electrons.

We assess the behavior of each of the downfolding methods upon doping by running the same calculations with a hole doping of $0.2$. We achieve this doping using the virutal crystal approximation, where we artificially change the Nd atomic number to $59.8$ and take out $0.2$ electrons from the system. We choose the number $0.2$ since that doping level is well within the experimental superconducting dome \cite{li2020superconducting}. 

\begin{table}[h]
\begin{tabular}{|c|c|c|}
\hline
               & $d_{x^2-y^2}$ & $d_{z^2}$ \\ \hline
MLWF           & -0.048        & -0.004    \\ \hline
SLWF           & -0.078        & -0.003    \\ \hline
Proj -10 to 10 & -0.053        & 0.004     \\ \hline
Proj -10 to 3  & -0.049        & 0.015     \\ \hline
\end{tabular}
\caption{Orbital occupanices (summed over spin) at hole doping of 0.20 minus the orbital occupanices at stoichiometry. Positive values indicate that the filling increases with hole doping.}
\label{tab:doping}
\end{table}

\begin{figure}[b]
    \centering
    \includegraphics[width = \linewidth]{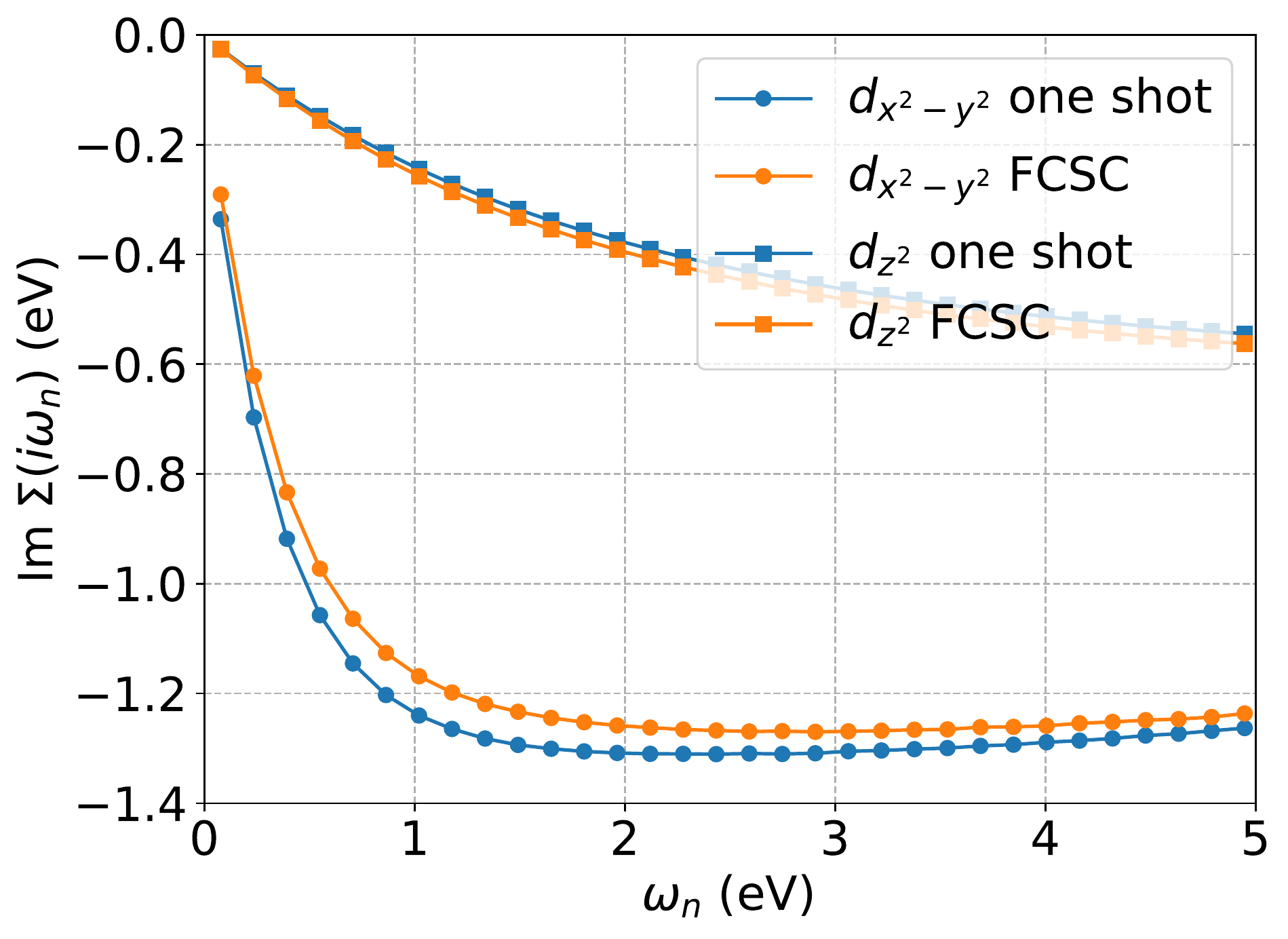}
    \caption{Comparison of the one shot and FCSC DFT+DMFT Imaginary Matsubara self energies for the case of projectors in the wide energy window from $-10$ to $\SI{10}{eV}$}
    \label{fig:fcsc_vs_one_shot_sigma}
\end{figure}

Table \ref{tab:doping} shows the changes in orbital fillings upon hole doping. For each method, the $d_{x^2-y^2}$ filling decreases with hole doping, as expected. The filling decreases more in the SLWF case than the other cases.  In all cases, the $d_{z^2}$ filling does not change considerably, but the behavior depends on the method. It increases the most in the projector from -10 to 3 case because the self doping band moves above $\SI{3}{eV}$ so more $d_{z^2}$ weight has to be given to the main $d_{z^2}$-derived band below the Fermi level. In the projector from -10 to 10 case it also increases. In the MLWF and SLWF cases it decreases. 

\section{Full Charge Self Consistency}

For the case of projectors in the wide window from $-10$ to $\SI{10}{eV}$, we compare the results of a fully charge self consistent (FCSC) DFT+DMFT calculation to the one-shot results. In the FCSC case, we find the same $d_{z^2}$ orbital occupancy as the one shot case and a $d_{x^2-y^2}$ occupancy of $1.16$, very close to the one shot results of $1.14$. Likewise, the self energies are not so different in the one shot and FCSC case, as shown in Figure \ref{fig:fcsc_vs_one_shot_sigma}. We can therefore conclude that full charge self consistency is not crucial to the DFT+DMFT study of \NNO{}.

\bibliography{references.bib}

\end{document}